\documentclass{article}
\usepackage{amsthm}
\usepackage{amsmath, amssymb, amsthm}  
\usepackage{fancyhdr}                  
\usepackage{graphicx}                  
\usepackage{float}
\usepackage{hyperref}                  
\usepackage{geometry}                  
\usepackage{natbib}
\usepackage{subcaption}
\usepackage{booktabs} 
\usepackage{longtable}
\usepackage{tikz}
\usepackage{array}
\usepackage{pdflscape}
\usepackage{caption}
\usepackage{color}
\title{Jhonata1}
\author{Carlos Diniz}
\date{October 2025}

\title{Matrix-Variate Regression Model for Multivariate Spatio-Temporal Data}

\author{
  Carlos A. R. Diniz, Victor E. Lachos \\ 
  \small{Statistics Department - Federal University of Sao Carlos, São Carlos, SP, Brazil} \\
  \and
  Victor H. Lachos \\ 
  \small{Statistics Department - University of Connecticut, Storrs, CT-06269, USA }
}

\begin{document}
\maketitle  

\begin{abstract}
This paper introduces a matrix-variate regression model for analyzing multivariate data observed across spatial locations and over time. The model's design incorporates a mean structure that links covariates to the response matrix and a separable covariance structure, based on a Kronecker product, to capture spatial and temporal dependencies efficiently. We derive maximum likelihood estimators for all model parameters. A simulation study validates the model, showing its effectiveness in parameter recovery across different spatial resolutions. Finally, an application to real-world data on agricultural and livestock production from Brazilian municipalities showcases the model's practical utility in revealing structured spatio-temporal patterns of variation and covariate effects.\\

\end{abstract}

\date{} 

\section{Introduction}

As complex, multi-dimensional datasets become increasingly common across scientific disciplines, there is a growing need for statistical models that move beyond traditional vector-valued approaches. In fields such as environmental science, neuroimaging, insurance analytics, and agricultural production, both responses and covariates are often naturally represented in matrix form, indexed by space, time, or experimental conditions. Modeling such data by vectorizing or isolating a single dimension typically leads to a loss of structure and interpretability, motivating regression frameworks that operate directly on matrices while capturing their inherent dependence patterns.

An important step toward this goal was taken by \citet{ding2018matrix}, who developed a general framework for matrix-variate regressions designed for data in which the response itself is a random matrix and the predictors may be scalar, vector, or matrix-valued. Their formulation explicitly accounts for correlations among elements of the response, avoiding the loss of information that arises from vectorization or separate modeling of rows and columns. Within this framework, the authors further introduced envelope extensions that eliminate immaterial variation and substantially reduce the number of parameters, leading to more efficient estimation in high-dimensional settings. This work established a key foundation for modeling matrix-valued responses while preserving their intrinsic dependence structure.

Building upon these ideas, \citet{viroli2021} formalized a general matrix-variate regression framework for three-way data, in which each observation is expressed as a response matrix rather than a vector. By assuming a matrix-variate normal structure for the error term, the model extends classical multiple and multivariate regression to explicitly capture dependencies along the two modes of the response. The separable covariance specification proposed by the author allows overall variability to be decomposed into within- and between-variable components, thereby enhancing interpretability across both dimensions of the data. Together, the contributions of \citet{ding2018matrix} and \citet{viroli2021} laid the groundwork for subsequent developments in matrix-variate regression, emphasizing interpretability through structured covariance formulations.

While these models provide an elegant framework for capturing two-dimensional dependencies, many real-world problems exhibit more complex dependence structures. In numerous contexts, multiple response variables are measured simultaneously across spatial locations and over time, giving rise to datasets with an inherently three-way organization. In such cases, the static matrix-variate formulation becomes insufficient, as additional layers of correlation—both temporal and spatial—must be incorporated to adequately describe the joint dynamics of the responses.

For instance, in environmental studies, a data matrix may represent climate variables (rows) measured across regions (columns) over multiple years (third mode). Likewise, in agricultural or biomedical experiments, several correlated outcomes are often recorded repeatedly across experimental units and time points. Analyzing each response separately or flattening the data into vectors risks overlooking meaningful interactions among variables, locations, and time periods. These challenges have motivated the development of dynamic and spatial extensions of matrix-variate models capable of accommodating such three-way structures.

An early step in this direction was provided by \citet{salvador2004analysis}, who developed a Bayesian framework for matrix-normal dynamic linear models with unknown, potentially time-varying covariance matrices. To address the model’s non-conjugacy, the authors implemented Gibbs sampling and demonstrated the flexibility of their approach through applications to industrial and financial data, providing a foundation for later developments in spatio-temporal modeling.

Further advances include the spatial matrix-variate regression of \citet{lan2019geostatistical}, which employs a spatial Wishart process to model positive-definite matrices in diffusion tensor imaging, and the matrix autoregressive framework of \citet{hsu2021matrix}, which efficiently captures spatio-temporal dependence in time-indexed matrix data. More recently, \citet{boyle2024matrix} illustrated the practical potential of matrix-variate regression through an application to insurance losses across regions and time periods. Together, these studies highlight the growing relevance of matrix-valued formulations for modeling multiway dependencies while maintaining the natural organization of the data.

Motivated by these developments, we propose a matrix-variate regression model designed for data indexed by spatial locations and time points. The mean structure is expressed as a function of covariates, providing direct interpretability of their effects on the multivariate response. One covariance matrix captures dependencies among response variables, while the other incorporates spatio-temporal structure through a Kronecker product formulation. We present an estimation procedure for all model parameters and demonstrate its implementation.

The remainder of the paper is organized as follows. Section~\ref{sec:preliminaries} introduces the notation and preliminary results. Section~\ref{sec:reg_model} presents the proposed model for spatio-temporal data, detailing spatial and temporal covariance structures in Subsection~\ref{subsec:cov_structures} and possible configurations for the coefficient matrix in Subsection~\ref{subsec:B_structures}. Section~\ref{sec:estimation} describes the likelihood-based estimation procedure. Section~\ref{resid} discusses residual analysis and model diagnostics. Section~\ref{sec:simulation} reports simulation results assessing estimator performance, and Section~\ref{sec:application} applies the model to agricultural and livestock production data from Brazilian municipalities. Finally, Section~\ref{sec:conclusion} concludes the paper and discusses perspectives for future research.

\section{Preliminaries}\label{sec:preliminaries}

Before introducing the proposed distribution, we establish the notation that will be used consistently throughout this paper. Random matrices of dimension $p \times q$ are denoted by $\mathbf{Y}$, with elements represented by $Y_{ij}$ for $i = 1, \dots, p$ and $j = 1, \dots, q$. Covariate matrices are denoted by $\mathbf{X}$. The operator $|\cdot|$ denotes the determinant when applied to square matrices and the absolute value for scalars; $\|\cdot\|$ indicates the matrix norm; $\operatorname{tr}(\cdot)$ represents the trace of a matrix; and $\otimes$ denotes the Kronecker product. As is standard in probability theory, random variables are represented by uppercase letters, their realizations by lowercase letters, and vectors and matrices by boldface characters.

The MVN distribution with mean matrix $\mathbf{M}$ and covariance matrices $\boldsymbol{\Sigma}$ and $\boldsymbol{\Psi}$ of dimensions $p \times p$ and $r \times r$, respectively, has probability density function (pdf) given by
\begin{equation}
\phi(\mathbf{Y}\mid\mathbf{M},\boldsymbol{\Sigma},\boldsymbol{\Psi}) = 
\frac{1}{(2\pi)^{pr/2}|\boldsymbol{\Psi}|^{p/2}|\boldsymbol{\Sigma}|^{r/2}} 
\exp\left( -\frac{1}{2} \operatorname{tr}\left[\boldsymbol{\Psi}^{-1}(\mathbf{Y} - \mathbf{M})^{\top} \boldsymbol{\Sigma}^{-1}(\mathbf{Y} - \mathbf{M})\right] \right).
\label{eq: MVN}
\end{equation}
We write this as $\mathbf{Y} \sim \mathcal{N}_{p\times r}(\mathbf{M}, \boldsymbol{\Sigma}, \boldsymbol{\Psi})$.  An equivalent definition specifies the MVN distribution as a special case of the multivariate normal distribution.
Specifically, 
\begin{equation}
	\mathbf{Y}\sim \mathcal{N}_{p\times q}(\mathbf{M}, \boldsymbol{\Sigma},\boldsymbol{\Psi)} \iff \text{vec}(\mathbf{Y}) \sim \mathcal{N}_{pq}(\boldsymbol{\mu}=\text{vec}(\mathbf{M}),\boldsymbol{\Lambda}=\boldsymbol{\Psi}\otimes\boldsymbol{\Sigma}).
 \label{eq: iff}
\end{equation}
where $\mathcal{N}_{pq}(\cdot)$ denotes the multivariate normal distribution with mean $\boldsymbol{\mu}$ and covariance matrix $\boldsymbol{\Lambda}$,

\section{Matrix-variate regression model for spatio-temporal data}\label{sec:reg_model}

We consider a matrix-variate regression model for a response matrix 
\( \mathbf{Y} \in \mathbb{R}^{p \times r} \), where each row corresponds to a distinct response variable and each column represents a spatio-temporal measurement, that is, a combination of a spatial location and a time point. Formally, we set \( r = L \times T \), where \( L \) and \( T \) denote the number of spatial locations and time points, respectively.

This organization, with response variables in the rows and spatio-temporal measurements in the columns, accommodates settings where multiple outcomes are observed at several locations and times. It generalizes the conventional format used in imaging or longitudinal studies, while enabling joint modeling of dependencies among response variables, spatial locations, and temporal observations.

This formulation produces a single observation matrix \( \mathbf{Y} \) that aggregates all measurements. Unlike traditional multivariate settings with multiple independent replicates, here the matrix \( \mathbf{Y} \) captures the full spatio-temporal structure of the data, with each column corresponding to a unique location–time pair. Because dependencies may arise both across variables and across spatio-temporal measurements, the covariance structure must be carefully specified to capture the complexity of the data.

To make this organization more explicit, let 
\( \mathbf{Y} = [\mathbf{Y}_1, \mathbf{Y}_2, \dots, \mathbf{Y}_r] \), 
where each column \( \mathbf{Y}_j \in \mathbb{R}^p \) represents the measurements of all \( p \) response variables at a specific location–time combination. An illustrative representation of \( \mathbf{Y} \) is given below, where the columns correspond to all \( L \times T \) combinations of spatial locations and time points:
\[
\mathbf{Y} =
\begin{bmatrix}
Y_{11} & Y_{12} & \hdots & Y_{1,LT} \\
Y_{21} & Y_{22} & \hdots & Y_{2,LT} \\
\vdots & \vdots & \ddots & \vdots \\
Y_{p1} & Y_{p2} & \hdots & Y_{p,LT}
\end{bmatrix}.
\]
In this format, the \( j \)th column of \( \mathbf{Y} \) contains the \( p \) responses observed at a particular combination of spatial location and time point. The index \( j \in \{1, \dots, LT\} \) can be mapped to a pair \( (\ell, t) \), with \( \ell \in \{1, \dots, L\} \) and \( t \in \{1, \dots, T\} \).

\medskip
We now introduce the statistical model adopted in this work. We assume that
\[
\mathbf{Y} \sim \mathcal{N}_{p \times r} \left( \mathbf{M}, \boldsymbol{\Sigma}, \boldsymbol{\Psi} \right),
\]
where:
\begin{itemize}
    \item \( \mathbf{M} \in \mathbb{R}^{p \times r} \) is the mean matrix;
    \item \( \boldsymbol{\Sigma} \in \mathbb{R}^{p \times p} \) is the row covariance matrix, describing dependencies among the response variables;
    \item \( \boldsymbol{\Psi} \in \mathbb{R}^{r \times r} \) is the column covariance matrix, capturing dependencies across space and time.
\end{itemize}

We adopt a separable structure for the column covariance matrix to represent spatial and temporal dependence through distinct components. Specifically,
\[
\boldsymbol{\Psi} = \boldsymbol{\Psi}_{\text{sp}} \otimes \boldsymbol{\Psi}_{\text{tp}},
\]
where \( \boldsymbol{\Psi}_{\text{sp}} \in \mathbb{R}^{L \times L} \) models spatial correlations 
(e.g., exponential or Matérn), and \( \boldsymbol{\Psi}_{\text{tp}} \in \mathbb{R}^{T \times T} \) 
models temporal dependence (e.g., autoregressive of order one, AR(1)). 
This separable representation simplifies the joint spatio-temporal dependence structure 
without implying independence between space and time.

\medskip
The mean structure is expressed as a linear function of a covariate matrix 
\( \mathbf{X} \in \mathbb{R}^{q \times r} \) through a coefficient matrix 
\( \mathbf{B} \in \mathbb{R}^{p \times q} \):
\[
\mathbf{M} = \mathbf{B}\mathbf{X},
\]
so that \( \mathbf{M} \) represents the mean matrix of the matrix-normal distribution. 
This formulation extends classical matrix-variate regression models by allowing 
\( p \) distinct response variables to be jointly modeled across spatial and temporal domains. 
It thus provides a unified framework for analyzing multivariate responses that evolve both over time 
and across locations.

\medskip
Summarizing, the complete model can be written as
\[
\mathbf{Y} = \mathbf{B}\mathbf{X} + \mathbf{E},
\]
where
\[
\mathbf{E} \sim \mathcal{N}_{p \times r}
\left( \mathbf{0},\, \boldsymbol{\Sigma},\, \boldsymbol{\Psi}_{\text{sp}} \otimes \boldsymbol{\Psi}_{\text{tp}} \right).
\]
Here, \( \mathbf{B} \) links the \( q \) covariates to the \( p \) responses, 
\( \boldsymbol{\Sigma} \) captures dependencies among responses, 
and \( \boldsymbol{\Psi}_{\text{sp}} \otimes \boldsymbol{\Psi}_{\text{tp}} \) 
encodes spatial and temporal correlations jointly. 
This compact expression makes explicit the dual dependence structure—across variables and across 
spatio-temporal measurements—that distinguishes matrix-variate regression models from 
standard multivariate or time-series formulations.

\medskip
Under this specification, \( \mathbb{E}[\mathbf{Y}] = \mathbf{B}\mathbf{X} \), 
so that \( \mathbb{E}[\mathbf{Y}] = \mathbf{0}_{p \times r} \) when \( \mathbf{X} = \mathbf{0}_{q \times r} \). 
This assumption is suitable when zero covariate values correspond to the absence of the phenomena under study. 
In other cases, a nonzero mean level may persist even when all covariates are zero, 
and an intercept term should be included. 
This can be achieved by augmenting \( \mathbf{X} \) with a row of ones, 
adding a corresponding column of constants in \( \mathbf{B} \). 
The extension is straightforward and fully compatible with the proposed framework, 
as it simply expands the dimensions of \( \mathbf{B} \) and \( \mathbf{X} \) 
without altering their internal structures or the predefined configurations discussed in 
Subsection~\ref{subsec:B_structures}. 
For simplicity, the following developments focus on the specification without an intercept.

\medskip
To complete the model specification, the next subsections detail the structures adopted for 
the covariance matrices and the coefficient matrix \( \mathbf{B} \). 
We first discuss the spatial and temporal covariance models used for 
\( \boldsymbol{\Psi}_{\text{sp}} \) and \( \boldsymbol{\Psi}_{\text{tp}} \), respectively, 
and then describe six alternative structures for \( \mathbf{B} \) reflecting 
different assumptions about the relationships between the covariates 
and the multivariate spatio-temporal response.

\subsection{Spatial and temporal covariance structures}\label{subsec:cov_structures}

In this subsection, we describe the parametric forms adopted for 
\( \boldsymbol{\Psi}_{\text{sp}} \) and \( \boldsymbol{\Psi}_{\text{tp}} \).
Throughout the paper, we consider {five} spatial covariance structures, Matérn, exponential, Gaussian, cubic, and spherical, and a {temporal} AR(1) structure. 
These choices provide a spectrum of smoothness and range behaviors (including compact support for cubic and spherical), while maintaining a common separable specification for the spatio–temporal covariance.

\subsubsection*{Matérn spatial covariance structure}
The Matérn covariance \citep{Cressie1993,matern_covariance} controls both range and smoothness:
\[
C(h)=\sigma_s^2 \frac{2^{1-\nu}}{\Gamma(\nu)}\!\left(\frac{\sqrt{2\nu}\,h}{\phi_s}\right)^{\!\nu}
K_\nu\!\left(\frac{\sqrt{2\nu}\,h}{\phi_s}\right),
\]
where $h$ is distance, $\sigma_s^2$ the variance, $\phi_s$ the range, and $\nu>0$ the smoothness (larger $\nu$ $\Rightarrow$ smoother fields). This family flexibly spans a wide spectrum of spatial regularity.

\subsubsection*{Exponential spatial covariance structure}
The exponential model \citep{Cressie1993} assumes a Markov property and yields rougher fields:
\[
C(h)=\sigma_s^2 \exp\!\left(-\frac{h}{\phi_s}\right).
\]
Here $\sigma_s^2$ is the variance and $\phi_s$ governs correlation decay. It is the Matérn case $\nu=0.5$.

\subsubsection*{Gaussian spatial covariance structure}\label{sec:gaussian-cov}
The Gaussian model \citep{Cressie1993} produces very smooth processes:
\[
C(h)=\sigma_s^2 \exp\!\left(-\frac{h^{2}}{\phi_s^{2}}\right).
\]
Compared to the exponential, decay is quadratic in the exponent; it is the Matérn limit as $\nu\to\infty$.

\subsubsection*{Cubic spatial covariance structure}
The cubic model \citep{Cressie1993} is compactly supported (zero beyond $2\phi_s$) and smoother than spherical:
\[
C(h)=\sigma_s^2\!\left[1-7\!\left(\frac{h}{2\phi_s}\right)^{\!2}
+\frac{35}{4}\!\left(\frac{h}{2\phi_s}\right)^{\!3}
-\frac{7}{2}\!\left(\frac{h}{2\phi_s}\right)^{\!5}
+\frac{3}{4}\!\left(\frac{h}{2\phi_s}\right)^{\!7}\right],\quad 0\le h\le 2\phi_s,
\]
and $C(h)=0$ for $h>2\phi_s$.

\subsubsection*{Spherical spatial covariance structure}
The spherical model \citep{Cressie1993} has a finite range $\phi_s$ (compact support):
\[
C(h)=
\begin{cases}
\sigma_s^2\!\left[1-\tfrac{3}{2}\tfrac{h}{\phi_s}+\tfrac{1}{2}\!\left(\tfrac{h}{\phi_s}\right)^{\!3}\right], & 0\le h\le \phi_s,\\[0.6em]
0, & h>\phi_s.
\end{cases}
\]
It features a gradual short‐range decay with an exact cutoff at $h=\phi_s$.

\subsubsection*{AR(1) temporal covariance structure}

Temporal dependence is modeled with a first-order autoregressive process (AR(1)) \citep{ar1_covariance}, where correlation decays geometrically with time separation:
\[
C(t_1,t_2)=\sigma_t^2 \rho^{\,|t_1-t_2|}, \quad |\rho|<1.
\]
Here $\sigma_t^2$ is the temporal variance and $\rho$ the autocorrelation parameter. The AR(1) structure captures stronger dependence at short lags and weaker correlation as time separation increases.

\subsection[\texorpdfstring{$\mathbf{B}$}{B}]{Possible structures for the coefficient matrix $\mathbf{B}$}\label{subsec:B_structures}

In the matrix-variate regression model, the coefficient matrix \( \mathbf{B} \in \mathbb{R}^{p \times q} \) plays a central role in linking the covariates to the multivariate spatio-temporal responses. The choice of structure for \( \mathbf{B} \) reflects modeling assumptions regarding how covariates influence the response variables and can significantly impact both interpretability and estimation efficiency.

In this subsection, we present a range of structural specifications for \( \mathbf{B} \), encompassing simple and low-dimensional forms as well as more flexible and complex configurations. These include identity and diagonal matrices, full (dense) matrices, interaction and polynomial expansions, sparse representations, and block matrix structures. Each alternative reflects a particular analytical goal or assumption about the underlying relationship between covariates and responses.

We now describe each of these structures in detail, highlighting their motivations, assumptions, and potential applications.

\subsubsection{Identity matrix.}
When \( p = q \) and \( \mathbf{B} = \mathbf{I}_p \), the identity matrix, each response row in \( \mathbf{Y} \) is explained directly by the corresponding row in \( \mathbf{X} \). This leads to a model in which each response variable depends on a single covariate, and all responses share a common error structure across columns. It is the most restrictive specification and results in a diagonal mapping from covariates to responses. While mathematically valid, this configuration is rarely useful in practice, as it imposes a rigid one-to-one relationship that prevents interactions or shared influences among covariates. Such a simplistic assumption seldom reflects the complexity found in real-world datasets.

\subsubsection{Diagonal matrix.}
When \( p = q \), meaning that the number of response variables matches the number of covariates, the coefficient matrix \( \mathbf{B} \) can be specified as diagonal. In this case, each response variable is influenced solely by a corresponding covariate, but with its own weight. This structure preserves a one-to-one relationship between covariates and responses while allowing for varying strengths of association. 

Consider the matrix regression model:
\[
\mathbf{Y} = \mathbf{B} \mathbf{X} + \mathbf{E},
\]
where
\[
\mathbf{B} =
\begin{bmatrix}
\beta_1 & 0 & 0 \\
0 & \beta_2 & 0 \\
0 & 0 & \beta_3
\end{bmatrix} \in \mathbb{R}^{3 \times 3}, \quad
\mathbf{X} =
\begin{bmatrix}
X_{11} & X_{12} \\
X_{21} & X_{22} \\
X_{31} & X_{32}
\end{bmatrix} \in \mathbb{R}^{3 \times 2},
\]
\[
\mathbf{Y} =
\begin{bmatrix}
Y_{11} & Y_{12} \\
Y_{21} & Y_{22} \\
Y_{31} & Y_{32}
\end{bmatrix} \in \mathbb{R}^{3 \times 2}, \quad
\mathbf{E} =
\begin{bmatrix}
\varepsilon_{11} & \varepsilon_{12} \\
\varepsilon_{21} & \varepsilon_{22} \\
\varepsilon_{31} & \varepsilon_{32}
\end{bmatrix} \in \mathbb{R}^{3 \times 2}.
\]

The product \( \mathbf{B} \mathbf{X} \) yields:
\[
\mathbf{B} \mathbf{X} =
\begin{bmatrix}
\beta_1 X_{11} & \beta_1 X_{12} \\
\beta_2 X_{21} & \beta_2 X_{22} \\
\beta_3 X_{31} & \beta_3 X_{32}
\end{bmatrix}.
\]

\subsubsection{Combination of covariates.}
To allow each response variable to be influenced by a combination of multiple covariates, the coefficient matrix \( \mathbf{B} \) must be dense. In this setting, each entry \( B_{ij} \) represents the contribution of the \(j \)th covariate to the \( i \)th response, enabling a flexible and expressive representation of the covariate-response relationships. This structure is widely adopted in classical multivariate regression, where each response may depend on all covariates to varying degrees.

In spatial or spatiotemporal contexts, this formulation allows the response at a given location to depend not only on covariates measured at that location but also on covariates from other locations. These cross-location effects are encoded in the off-diagonal entries of the coefficient matrix \( \mathbf{B} \), enabling the model to account for spatial interactions or spillover effects across units.

Consider the matrix regression model:
\[
\mathbf{Y} = \mathbf{B} \mathbf{X} + \mathbf{E},
\]
where
\[
\mathbf{B} =
\begin{bmatrix}
\beta_{11} & \beta_{12} & \beta_{13} \\
\beta_{21} & \beta_{22} & \beta_{23} \\
\beta_{31} & \beta_{32} & \beta_{33}
\end{bmatrix} \in \mathbb{R}^{3 \times 3}, \quad
\mathbf{X} =
\begin{bmatrix}
X_{11} & X_{12} \\
X_{21} & X_{22} \\
X_{31} & X_{32}
\end{bmatrix} \in \mathbb{R}^{3 \times 2},
\]
\[
\mathbf{Y} =
\begin{bmatrix}
Y_{11} & Y_{12} \\
Y_{21} & Y_{22} \\
Y_{31} & Y_{32}
\end{bmatrix} \in \mathbb{R}^{3 \times 2}, \quad
\mathbf{E} =
\begin{bmatrix}
\varepsilon_{11} & \varepsilon_{12} \\
\varepsilon_{21} & \varepsilon_{22} \\
\varepsilon_{31} & \varepsilon_{32}
\end{bmatrix} \in \mathbb{R}^{3 \times 2}.
\]

The matrix product \( \mathbf{B} \mathbf{X} \) is computed as:
\[
\mathbf{B} \mathbf{X} =
\begin{bmatrix}
\beta_{11} X_{11} + \beta_{12} X_{21} + \beta_{13} X_{31} & \beta_{11} X_{12} + \beta_{12} X_{22} + \beta_{13} X_{32} \\
\beta_{21} X_{11} + \beta_{22} X_{21} + \beta_{23} X_{31} & \beta_{21} X_{12} + \beta_{22} X_{22} + \beta_{23} X_{32} \\
\beta_{31} X_{11} + \beta_{32} X_{21} + \beta_{33} X_{31} & \beta_{31} X_{12} + \beta_{32} X_{22} + \beta_{33} X_{32}
\end{bmatrix}.
\]

In this formulation, each element of \( \mathbf{Y} \) is expressed as a linear combination of covariates from all locations, weighted by the corresponding coefficients in \( \mathbf{B} \), and perturbed by additive noise in \( \mathbf{E} \). For example, the response at the first location depends not only on covariates from that location but also on covariates from the second and third locations, reflecting the model’s capacity to capture contextual and spatial dependencies.

\subsubsection{Inclusion of interaction and polynomial terms.}
Interaction effects and nonlinearities can be incorporated by augmenting the covariate matrix \( \mathbf{X} \) to include pairwise interactions and polynomial expansions such as squared or cubic terms. The resulting matrix \( \widetilde{\mathbf{X}} \) consists of both the original covariates and these derived terms, while the corresponding coefficient matrix \( \widetilde{\mathbf{B}} \) encodes their effects on the response.

This extension enhances the model's flexibility and enables it to capture more complex relationships. However, it also increases the number of parameters, raising the risk of overfitting—particularly when the sample size is limited. To mitigate this issue, regularization techniques can be employed. For example, Lasso regularization imposes an \(\ell_1\) penalty on the coefficients, encouraging sparsity in \( \widetilde{\mathbf{B}} \) by shrinking small effects to zero and improving model generalization.

For illustration, suppose the original covariate matrix is:
\[
\mathbf{X} =
\begin{bmatrix}
X_{11} & X_{12} \\
X_{21} & X_{22} \\
X_{31} & X_{32}
\end{bmatrix} \in \mathbb{R}^{3 \times 2},
\]
containing two covariates across three locations. An augmented version with interaction and squared terms becomes:
\[
\widetilde{\mathbf{X}} =
\begin{bmatrix}
X_{11} & X_{12} & X_{11} X_{12} & X_{11}^2 & X_{12}^2 \\
X_{21} & X_{22} & X_{21} X_{22} & X_{21}^2 & X_{22}^2 \\
X_{31} & X_{32} & X_{31} X_{32} & X_{31}^2 & X_{32}^2
\end{bmatrix} \in \mathbb{R}^{3 \times 5}.
\]

Each row contains the main effects, interaction, and quadratic terms. The associated coefficient matrix \( \widetilde{\mathbf{B}} \in \mathbb{R}^{3 \times 5} \) is defined as:
\[
\widetilde{\mathbf{B}} =
\begin{bmatrix}
\beta_{11} & \beta_{12} & \beta_{13} & \beta_{14} & \beta_{15} \\
\beta_{21} & \beta_{22} & \beta_{23} & \beta_{24} & \beta_{25} \\
\beta_{31} & \beta_{32} & \beta_{33} & \beta_{34} & \beta_{35}
\end{bmatrix}.
\]

The matrix regression model then becomes:
\[
\mathbf{Y} = \widetilde{\mathbf{B}} \widetilde{\mathbf{X}} + \mathbf{E},
\]
which allows each response to be modeled as a flexible nonlinear combination of the covariates.

Explicitly, the entries of the response matrix \( \mathbf{Y} = [Y_{ij}] \in \mathbb{R}^{3 \times 2} \) can be written as:
\[
\begin{aligned}
Y_{1j} &= \beta_{11} X_{1j} + \beta_{12} X_{1,j+1} + \beta_{13} X_{1j} X_{1,j+1} + \beta_{14} X_{1j}^2 + \beta_{15} X_{1,j+1}^2 + \varepsilon_{1j}, \\
Y_{2j} &= \beta_{21} X_{2j} + \beta_{22} X_{2,j+1} + \beta_{23} X_{2j} X_{2,j+1} + \beta_{24} X_{2j}^2 + \beta_{25} X_{2,j+1}^2 + \varepsilon_{2j}, \\
Y_{3j} &= \beta_{31} X_{3j} + \beta_{32} X_{3,j+1} + \beta_{33} X_{3j} X_{3,j+1} + \beta_{34} X_{3j}^2 + \beta_{35} X_{3,j+1}^2 + \varepsilon_{3j},
\end{aligned}
\]
for \( j = 1, 2 \), where \( X_{ij} \) and \( X_{i,j+1} \) refer to the first and second covariates for the \( i \)th location, and \( \varepsilon_{ij} \) are the respective error terms.

This formulation makes it clear how each response incorporates linear, interaction, and nonlinear (polynomial) effects of the covariates. It provides flexibility for modeling complex relationships, though at the cost of increased dimensionality and potential overfitting.

\subsubsection{Sparse structure}

In high-dimensional settings, where the number of covariates (or derived terms) may exceed the number of observations, it is often desirable to impose sparsity on the coefficient matrix \( \mathbf{B} \). Sparsity means that many entries of \( \mathbf{B} \) are constrained to be exactly zero, effectively performing covariate selection within the matrix-variate regression framework.

Sparsity can be induced through regularization techniques such as Lasso or group Lasso, which penalize the absolute size of the coefficients and encourage many of them to shrink to zero. Alternatively, sparsity may be imposed based on prior knowledge or subject-matter expertise. In such cases, the researcher specifies a sparse structure for \( \mathbf{B} \), guided by theoretical reasoning, empirical findings, or contextual understanding of the covariate-response relationships.

\medskip
\noindent
As an illustration, consider the following regression model:
\[
\mathbf{Y} = \mathbf{B} \mathbf{X} + \mathbf{E},
\]
with \( \mathbf{B}, \mathbf{X}, \mathbf{Y}, \mathbf{E} \in \mathbb{R}^{3 \times 3} \). Suppose the coefficient matrix \( \mathbf{B} \) is sparse, with the structure:
\[
\mathbf{B} =
\begin{bmatrix}
\beta_{11} & 0 & \beta_{13} \\
0 & \beta_{22} & 0 \\
\beta_{31} & 0 & 0
\end{bmatrix},
\quad
\mathbf{X} =
\begin{bmatrix}
X_{11} & X_{12} & X_{13} \\
X_{21} & X_{22} & X_{23} \\
X_{31} & X_{32} & X_{33}
\end{bmatrix}.
\]

The matrix product \( \mathbf{B} \mathbf{X} \) becomes:
\[
\begin{bmatrix}
\beta_{11} X_{11} + \beta_{13} X_{31} & \beta_{11} X_{12} + \beta_{13} X_{32} & \beta_{11} X_{13} + \beta_{13} X_{33} \\
\beta_{22} X_{21} & \beta_{22} X_{22} & \beta_{22} X_{23} \\
\beta_{31} X_{11} & \beta_{31} X_{12} & \beta_{31} X_{13}
\end{bmatrix}.
\]

Thus, the full regression model becomes:
\[
\mathbf{Y} =
\begin{bmatrix}
\beta_{11} X_{11} + \beta_{13} X_{31} + \varepsilon_{11} &
\beta_{11} X_{12} + \beta_{13} X_{32} + \varepsilon_{12} &
\beta_{11} X_{13} + \beta_{13} X_{33} + \varepsilon_{13} \\
\beta_{22} X_{21} + \varepsilon_{21} &
\beta_{22} X_{22} + \varepsilon_{22} &
\beta_{22} X_{23} + \varepsilon_{23} \\
\beta_{31} X_{11} + \varepsilon_{31} &
\beta_{31} X_{12} + \varepsilon_{32} &
\beta_{31} X_{13} + \varepsilon_{33}
\end{bmatrix}.
\]

This representation clearly shows how sparsity in \( \mathbf{B} \) allows for selective covariate-response relationships, improving both interpretability and estimation efficiency, particularly in high-dimensional or structured regression settings.

\subsubsection{Block matrix structure}

In matrix-variate regression models, where the mean matrix is specified as \( \mathbf{M} = \mathbf{B} \mathbf{X} \), it is often useful to consider a block structure for the coefficient matrix \( \mathbf{B} \). This arises naturally when both covariates and responses can be grouped according to spatial, temporal, or functional characteristics. Each block in \( \mathbf{B} \) captures the influence of a group of covariates on a group of responses, allowing for modular modeling strategies such as sparsity or low-rank constraints within blocks.

\medskip
\noindent
As an illustrative example, suppose we have four response variables and three covariates, organized into two groups: responses 1 and 2 depend on covariates 1 and 2, and responses 3 and 4 depend only on covariate 3. This leads to a block structure in the coefficient matrix \( \mathbf{B} \in \mathbb{R}^{4 \times 3} \):
\[
\mathbf{B} =
\begin{bmatrix}
\beta_{11} & \beta_{12} & 0 \\
\beta_{21} & \beta_{22} & 0 \\
0 & 0 & \beta_{33} \\
0 & 0 & \beta_{43}  
\end{bmatrix}.
\]

We now consider the full matrix regression model:
\[
\mathbf{Y} = \mathbf{B} \mathbf{X} + \mathbf{E},
\]
where
\[
\mathbf{X} =
\begin{bmatrix}
X_{11} & X_{12} & X_{13} \\
X_{21} & X_{22} & X_{23} \\
X_{31} & X_{32} & X_{33}
\end{bmatrix} \in \mathbb{R}^{3 \times 3}, \quad
\mathbf{Y}, \mathbf{E} \in \mathbb{R}^{4 \times 3}.
\]

The matrix product \( \mathbf{B} \mathbf{X} \) becomes:
\[
\mathbf{B} \mathbf{X} =
\begin{bmatrix}
\beta_{11} X_{11} + \beta_{12} X_{21} & \beta_{11} X_{12} + \beta_{12} X_{22} & \beta_{11} X_{13} + \beta_{12} X_{23} \\
\beta_{21} X_{11} + \beta_{22} X_{21} & \beta_{21} X_{12} + \beta_{22} X_{22} & \beta_{21} X_{13} + \beta_{22} X_{23} \\
\beta_{33} X_{31} & \beta_{33} X_{32} & \beta_{33} X_{33} \\
\beta_{43} X_{31} & \beta_{43} X_{32} & \beta_{43} X_{33}
\end{bmatrix}.
\]

Therefore, the full regression model is:
\[
\mathbf{Y} =
\begin{bmatrix}
\beta_{11} X_{11} + \beta_{12} X_{21} + \varepsilon_{11} & \beta_{11} X_{12} + \beta_{12} X_{22} + \varepsilon_{12} & \beta_{11} X_{13} + \beta_{12} X_{23} + \varepsilon_{13} \\
\beta_{21} X_{11} + \beta_{22} X_{21} + \varepsilon_{21} & \beta_{21} X_{12} + \beta_{22} X_{22} + \varepsilon_{22} & \beta_{21} X_{13} + \beta_{22} X_{23} + \varepsilon_{23} \\
\beta_{33} X_{31} + \varepsilon_{31} & \beta_{33} X_{32} + \varepsilon_{32} & \beta_{33} X_{33} + \varepsilon_{33} \\
\beta_{43} X_{31} + \varepsilon_{41} & \beta_{43} X_{32} + \varepsilon_{42} & \beta_{43} X_{33} + \varepsilon_{43}
\end{bmatrix}.
\]

This formulation shows how block structures in \( \mathbf{B} \) can reflect group-wise dependencies and reduce model complexity while maintaining interpretability.

\section{Matrix-variate regression model and likelihood-based estimation}\label{sec:estimation}

Let \( \mathbf{Y} \in \mathbb{R}^{p \times r} \) denote a matrix-variate observation from the model
\[
\mathbf{Y} \sim \mathcal{N}_{p \times r}(\mathbf{M}, \boldsymbol{\Sigma}, \boldsymbol{\Psi}),
\]
where \( \mathbf{M} = \mathbf{B} \mathbf{X} \in \mathbb{R}^{p \times r} \) is the mean matrix, constructed from a coefficient matrix \( \mathbf{B} \in \mathbb{R}^{p \times q} \) and a covariate matrix \( \mathbf{X} \in \mathbb{R}^{q \times r} \);  
\( \boldsymbol{\Sigma} \in \mathbb{R}^{p \times p} \) is the row covariance matrix, modeling dependencies across the response variables;  
and \( \boldsymbol{\Psi} \in \mathbb{R}^{r \times r} \) is the column covariance matrix, representing spatio-temporal dependence. 

Following the separable structure adopted in this work, the column covariance matrix is written as
\[
\boldsymbol{\Psi} = \boldsymbol{\Psi}_{\text{sp}} \otimes \boldsymbol{\Psi}_{\text{tp}},
\]
where \( \boldsymbol{\Psi}_{\text{sp}} \in \mathbb{R}^{L \times L} \) captures spatial dependence and \( \boldsymbol{\Psi}_{\text{tp}} \in \mathbb{R}^{T \times T} \) accounts for temporal dependence.

For the spatial component, \( \boldsymbol{\Psi}_{\text{sp}} \), we employ five alternative covariance structures: the Matérn, exponential, Gaussian, cubic, and spherical models. Each imparts a distinct degree of smoothness and correlation decay, allowing for flexible modeling of spatial dependence. For the temporal component, \( \boldsymbol{\Psi}_{\text{tp}} \), we assume a first-order autoregressive (AR(1)) structure to capture serial dependence. This combination yields a flexible yet interpretable spatio-temporal covariance matrix, capable of accommodating a wide range of dependence patterns.

Given these definitions, the log-likelihood function for the parameters \( (\mathbf{B}, \boldsymbol{\Sigma}, \boldsymbol{\Psi}) \), up to an additive constant, is given by
\[
\ell(\mathbf{B}, \boldsymbol{\Sigma}, \boldsymbol{\Psi}) =
- \frac{r}{2} \log |\boldsymbol{\Sigma}|
- \frac{p}{2} \log |\boldsymbol{\Psi}|
- \frac{1}{2} \mathrm{tr} \left[
\boldsymbol{\Sigma}^{-1} (\mathbf{Y} - \mathbf{B} \mathbf{X}) \boldsymbol{\Psi}^{-1} (\mathbf{Y} - \mathbf{B} \mathbf{X})^\top
\right].
\]

The first two terms originate from the normalization constants of the matrix-normal distribution. The third term represents the squared Mahalanobis distance between the observed data \( \mathbf{Y} \) and the mean structure \( \mathbf{M} = \mathbf{B} \mathbf{X} \), where the distance is evaluated using the inverse of the Kronecker covariance structure \( \boldsymbol{\Sigma} \otimes \boldsymbol{\Psi} \).

This likelihood function, however, is not identifiable: for any positive scalar \( a > 0 \), the transformation \( (\boldsymbol{\Sigma}, \boldsymbol{\Psi}) \mapsto (a \boldsymbol{\Sigma}, a^{-1} \boldsymbol{\Psi}) \) leaves the log-likelihood unchanged. To resolve this identifiability issue, it is customary to impose a constraint on one of the covariance matrices, such as fixing its trace, determinant, or a specific entry. In our implementation, we adopt the convention \( \Sigma_{1,1} = 1 \) to ensure identifiability and simplify estimation.

\subsection[\texorpdfstring{$\mathbf{B}$}{B}]{Estimation of $\mathbf{B}$}

The log-likelihood function (up to an additive constant) is given by:
\[
\ell(\mathbf{B}) = -\frac{1}{2} \, \mathrm{tr} \left[ \boldsymbol{\Sigma}^{-1} (\mathbf{Y} - \mathbf{B} \mathbf{X}) \boldsymbol{\Psi}^{-1} (\mathbf{Y} - \mathbf{B} \mathbf{X})^\top \right].
\]
Let us define the residual matrix \( \mathbf{E} = \mathbf{Y} - \mathbf{B} \mathbf{X} \). We denote the entries of the data matrices by \( \mathbf{Y} = [y_{ik}] \) and \( \mathbf{X} = [x_{ik}] \), where \( y_{ik} \) and \( x_{ik} \) represent the \((i,k)\)th elements of the response and covariate matrices, respectively.

\subsubsection[\texorpdfstring{$\mathbf{B}$}{B} diagonal]{Estimation under $\mathbf{B}$ diagonal}

Assuming that \( \mathbf{B} \) is diagonal, i.e., \( \mathbf{B} = \operatorname{diag}(\beta_{11}, \dots, \beta_{pp}) \), the product \( \mathbf{B} \mathbf{X} \) yields entries \( (\mathbf{B} \mathbf{X})_{ik} = \beta_{ii} x_{ik} \). Consequently, the residual matrix \( \mathbf{E} \) has elements \( e_{ik} = y_{ik} - \beta_{ii} x_{ik} \).

Let \( \boldsymbol{\Sigma}^{-1} = [\sigma^{ij}] \in \mathbb{R}^{p \times p} \) and \( \boldsymbol{\Psi}^{-1} = [\psi^{k\ell}] \in \mathbb{R}^{r \times r} \). Then the objective function becomes
\[
Q(\mathbf{B}) = \mathrm{tr} \left( \boldsymbol{\Sigma}^{-1} \mathbf{E} \boldsymbol{\Psi}^{-1} \mathbf{E}^\top \right),
\]
which, when expanded in terms of matrix elements, takes the form
\[
Q(\mathbf{B}) = \sum_{i=1}^{p} \sum_{j=1}^{p} \sum_{k=1}^{r} \sum_{\ell=1}^{r}
\sigma^{ij} \psi^{k\ell} \, e_{ik} e_{j\ell}.
\]
Substituting \( e_{ik} = y_{ik} - \beta_{ii} x_{ik} \), we obtain
\[
Q(\mathbf{B}) =
\sum_{i=1}^{p} \sum_{j=1}^{p} \sum_{k=1}^{r} \sum_{\ell=1}^{r}
\sigma^{ij} \psi^{k\ell}
\left( y_{ik} - \beta_{ii} x_{ik} \right)
\left( y_{j\ell} - \beta_{jj} x_{j\ell} \right).
\]

To derive the estimating equations, we compute the partial derivative of \( Q(\mathbf{B}) \) with respect to \( \beta_{tt} \):
\[
\frac{\partial Q}{\partial \beta_{tt}} = \sum_{i=1}^p \sum_{j=1}^p \sum_{k=1}^r \sum_{\ell=1}^r \sigma^{ij} \psi^{k\ell} \left( \frac{\partial e_{ik}}{\partial \beta_{tt}} e_{j\ell} + e_{ik} \frac{\partial e_{j\ell}}{\partial \beta_{tt}} \right).
\]

Note that
\[
\frac{\partial e_{ik}}{\partial \beta_{tt}} = 
\begin{cases}
- x_{ik}, & \text{if } i = t \\
0, & \text{otherwise}
\end{cases}
\quad \text{and} \quad
\frac{\partial e_{j\ell}}{\partial \beta_{tt}} = 
\begin{cases}
- x_{j\ell}, & \text{if } j = t \\
0, & \text{otherwise}.
\end{cases}
\]

Thus, the derivative simplifies to:
\[
\frac{\partial Q}{\partial \beta_{tt}} = -2 \sum_{j=1}^p \sum_{k=1}^r \sum_{\ell=1}^r \sigma^{tj} \psi^{k\ell} x_{tk} \left( y_{j\ell} - \beta_{jj} x_{j\ell} \right).
\]

Setting \( \frac{\partial Q}{\partial \beta_{tt}} = 0 \) gives the estimating equation:
\[
\sum_{j=1}^p \sum_{k=1}^r \sum_{\ell=1}^r \sigma^{tj} \psi^{k\ell} x_{tk} \left( y_{j\ell} - \beta_{jj} x_{j\ell} \right) = 0.
\]

Separating the \( j = t \) term from \( j \neq t \) terms:
\[
\sum_{k=1}^r \sum_{\ell=1}^r \sigma^{tt} \psi^{k\ell} x_{tk} x_{t\ell} \beta_{tt} + \sum_{\substack{j=1 \\ j \neq t}}^p \sum_{k=1}^r \sum_{\ell=1}^r \sigma^{tj} \psi^{k\ell} x_{tk} \left( y_{j\ell} - \beta_{jj} x_{j\ell} \right) = \sum_{k=1}^r \sum_{\ell=1}^r \sigma^{tt} \psi^{k\ell} x_{tk} y_{t\ell}.
\]

Solving for \( \beta_{tt} \), we obtain the explicit estimator:
\[
\hat{\beta}_{tt} = \frac{
\displaystyle
\sum_{k=1}^r \sum_{\ell=1}^r \hat{\sigma}^{tt} \hat{\psi}^{k\ell} x_{tk} y_{t\ell}
- \sum_{\substack{j=1 \\ j \neq t}}^p \sum_{k=1}^r \sum_{\ell=1}^r \hat{\sigma}^{tj} \hat{\psi}^{k\ell} x_{tk} \left( y_{j\ell} - \hat{\beta}_{jj} x_{j\ell} \right)
}{
\displaystyle
\sum_{k=1}^r \sum_{\ell=1}^r \hat{\sigma}^{tt} \hat{\psi}^{k\ell} x_{tk} x_{t\ell}
}.
\]

This formulation reveals that the diagonal coefficients are coupled through the off-diagonal elements of \( \boldsymbol{\Sigma}^{-1} \). When \( \boldsymbol{\Sigma} \) is also diagonal, the estimator simplifies to the standard weighted least squares solution for each response variable separately.

\subsubsection{Estimation under the combination of covariates structure}

We now consider the estimation of \( \mathbf{B} \) under the general case where each response may depend on a linear combination of covariates, allowing \( \mathbf{B} \in \mathbb{R}^{p \times q} \) to be a fully dense matrix.

The log-likelihood function (up to a constant) is given by
\[
\ell(\mathbf{B}) = -\frac{1}{2} \, \mathrm{tr}\left[ \boldsymbol{\Sigma}^{-1} (\mathbf{Y} - \mathbf{B} \mathbf{X}) \boldsymbol{\Psi}^{-1} (\mathbf{Y} - \mathbf{B} \mathbf{X})^\top \right].
\]

To obtain the maximum likelihood (ML) estimator of \( \mathbf{B} \), we differentiate \( \ell(\mathbf{B}) \) with respect to \( \mathbf{B} \) and set the result equal to zero. Using standard results from matrix calculus for the trace operator, the gradient is
\[
\frac{\partial \ell(\mathbf{B})}{\partial \mathbf{B}} = \boldsymbol{\Sigma}^{-1} (\mathbf{Y} - \mathbf{B} \mathbf{X}) \boldsymbol{\Psi}^{-1} \mathbf{X}^\top.
\]

Setting the derivative equal to zero leads to the estimating equation
\[
\boldsymbol{\Sigma}^{-1} (\mathbf{Y} - \mathbf{B} \mathbf{X}) \boldsymbol{\Psi}^{-1} \mathbf{X}^\top = \mathbf{0},
\]
which simplifies to
\[
\mathbf{Y} \hat{\boldsymbol{\Psi}}^{-1} \mathbf{X}^\top = \hat{\mathbf{B}} \mathbf{X} \hat{\boldsymbol{\Psi}}^{-1} \mathbf{X}^\top.
\]

Assuming that \( \mathbf{X} \hat{\boldsymbol{\Psi}}^{-1} \mathbf{X}^\top \) is invertible, we obtain the ML estimator:
\[
\hat{\mathbf{B}} = \mathbf{Y} \hat{\boldsymbol{\Psi}}^{-1} \mathbf{X}^\top \left( \mathbf{X} \hat{\boldsymbol{\Psi}}^{-1} \mathbf{X}^\top \right)^{-1}.
\]

In the case where interaction effects and nonlinearities, such as squared or higher-order polynomial terms, are included, the covariate matrix is expanded to \( \widetilde{\mathbf{X}} \), which incorporates both the original covariates and the additional derived terms. The estimation procedure remains analogous to the previous case, with the only adjustment being the consistent use of the augmented matrix \( \widetilde{\mathbf{X}} \) and the corresponding coefficient matrix \( \widetilde{\mathbf{B}} \).

\subsubsection{Estimation under sparse structure}

Since no systematic structure is assumed in this case, estimation must be carried out coefficient-wise. A natural strategy is to expand the log-likelihood function into scalar terms and compute partial derivatives with respect to each free parameter $\beta_{ij}$, excluding those fixed at zero. This leads to a system of equations that can be solved jointly to obtain the estimates, while respecting the predefined sparsity pattern of $\mathbf{B}$.

To formalize this coefficient-wise approach within the matrix-variate framework, we consider the constrained optimization problem where the coefficient matrix $\mathbf{B}$ exhibits a known sparse structure, requiring maximization of the log-likelihood function subject to sparsity constraints:

\[
\ell(\mathbf{B}) = -\frac{1}{2} \, \mathrm{tr} \left[ \boldsymbol{\Sigma}^{-1} (\mathbf{Y} - \mathbf{B} \mathbf{X}) \boldsymbol{\Psi}^{-1} (\mathbf{Y} - \mathbf{B} \mathbf{X})^\top \right],
\]
with $\beta_{ij} = 0$ for all $(i,j) \notin \mathcal{S}$, where $\mathcal{S}$ denotes the set of free parameter indices.

The partial derivatives with respect to free parameters yield the estimating equations:

\[
\frac{\partial \ell}{\partial \beta_{uv}} = \sum_{i=1}^p \sigma^{iu} \sum_{k=1}^r \sum_{l=1}^r \psi^{kl} e_{ik} x_{vl} = 0, \quad \forall (u,v) \in \mathcal{S},
\]
where $e_{ik} = y_{ik} - \sum_{m=1}^q \beta_{im} x_{mk}$ are the residual elements.

In matrix form, these conditions become:

\[
\left[ \boldsymbol{\Sigma}^{-1} (\mathbf{Y} - \mathbf{B} \mathbf{X}) \boldsymbol{\Psi}^{-1} \mathbf{X}^\top \right]_{uv} = 0, \quad \forall (u,v) \in \mathcal{S}.
\]

Let $\boldsymbol{\beta}_{\text{free}}$ be the vector of free parameters and $\mathbf{S}$ the selection matrix satisfying:

\[
\operatorname{vec}(\mathbf{B}) = \mathbf{S} \boldsymbol{\beta}_{\text{free}}.
\]

The estimating equations become:

\[
\mathbf{S}^\top \operatorname{vec}\left( \boldsymbol{\Sigma}^{-1} (\mathbf{Y} - \mathbf{B} \mathbf{X}) \boldsymbol{\Psi}^{-1} \mathbf{X}^\top \right) = \mathbf{0}.
\]

This leads to the linear system:

\[
\mathbf{H} \boldsymbol{\beta}_{\text{free}} = \mathbf{g},
\]
where:
\[ 
\mathbf{H} = \mathbf{S}^\top \left( \mathbf{X} \boldsymbol{\Psi}^{-1} \mathbf{X}^\top \otimes \boldsymbol{\Sigma}^{-1} \right)  \mathbf{S}  \quad \text{and} \quad
\mathbf{g} = \mathbf{S}^\top \operatorname{vec}\left( \boldsymbol{\Sigma}^{-1} \mathbf{Y} \boldsymbol{\Psi}^{-1} \mathbf{X}^\top \right). 
\]

To ensure numerical stability, we employ ridge regularization:

\[
\hat{\boldsymbol{\beta}}_{\text{free}} = (\hat{\mathbf{H}} + \lambda \mathbf{I})^{-1} \hat{\mathbf{g}},
\]
with $\lambda > 0$ chosen sufficiently small to maintain estimation accuracy while preventing ill-conditioning.

\subsubsection{Estimation under block-structured coefficient matrices}

We now consider the case where the coefficient matrix \( \mathbf{B} \) follows a block-diagonal structure, reflecting independent relationships between groups of covariates and groups of responses.

We assume the mean structure \( \mathbf{M} = \mathbf{B} \mathbf{X} \), where the coefficient matrix \( \mathbf{B} \in \mathbb{R}^{p \times q} \) is partitioned as
\[
\mathbf{B} =
\begin{bmatrix}
\mathbf{B}_1 & \mathbf{0} \\
\mathbf{0} & \mathbf{B}_2 
\end{bmatrix},
\quad \text{with} \quad
\mathbf{B}_1 \in \mathbb{R}^{p_1 \times q_1}, \quad
\mathbf{B}_2 \in \mathbb{R}^{p_2 \times q_2}, \quad
p_1 + p_2 = p, \quad q_1 + q_2 = q.
\]

Here, \( \mathbf{B}_1 \) models the effect of the first group of covariates on the first group of responses, and \( \mathbf{B}_2 \) models the second group analogously. The zero blocks indicate the absence of interaction across blocks. The covariate matrix \( \mathbf{X} \in \mathbb{R}^{q \times r} \) and the response matrix \( \mathbf{Y} \in \mathbb{R}^{p \times r} \) are partitioned conformably with the dimensions of \( \mathbf{B} \), that is,
\[
\mathbf{X} =
\begin{bmatrix}
\mathbf{X}_1 \\
\mathbf{X}_2
\end{bmatrix},
\qquad
\mathbf{Y} =
\begin{bmatrix}
\mathbf{Y}_1 \\
\mathbf{Y}_2
\end{bmatrix},
\]
where \( \mathbf{X}_1 \in \mathbb{R}^{q_1 \times r} \), \( \mathbf{X}_2 \in \mathbb{R}^{q_2 \times r} \), \( \mathbf{Y}_1 \in \mathbb{R}^{p_1 \times r} \), and \( \mathbf{Y}_2 \in \mathbb{R}^{p_2 \times r} \). Note that no zero structure is imposed on \( \mathbf{X} \) or \( \mathbf{Y} \); the partitioning is solely for consistency with the block structure of \( \mathbf{B} \).

Assuming independence between blocks and a shared column covariance \( \boldsymbol{\Psi} \in \mathbb{R}^{r \times r} \), the log-likelihood function for the block-structured model is given (up to a constant) by:
\[
\ell(\mathbf{B}_1, \mathbf{B}_2) =
- \frac{1}{2} \, \mathrm{tr} \left[
\boldsymbol{\Sigma}_1^{-1} (\mathbf{Y}_1 - \mathbf{B}_1 \mathbf{X}_1) \boldsymbol{\Psi}^{-1} (\mathbf{Y}_1 - \mathbf{B}_1 \mathbf{X}_1)^\top
\right]
-
\frac{1}{2} \, \mathrm{tr} \left[
\boldsymbol{\Sigma}_2^{-1} (\mathbf{Y}_2 - \mathbf{B}_2 \mathbf{X}_2) \boldsymbol{\Psi}^{-1} (\mathbf{Y}_2 - \mathbf{B}_2 \mathbf{X}_2)^\top
\right].
\]

This expression highlights that the log-likelihood naturally decomposes across blocks, allowing for separate estimation of \( \mathbf{B}_1 \) and \( \mathbf{B}_2 \).

Since the log-likelihood function decomposes additively across the blocks, the estimation of \( \mathbf{B}_1 \) and \( \mathbf{B}_2 \) reduces to two independent problems of the same form as in the fully dense case. Therefore, the ML estimators for each block are obtained using the same formula, applied separately to each pair \( (\mathbf{Y}_j, \mathbf{X}_j) \), for \( j = 1, 2 \).

The ML estimators for \( \mathbf{B}_1 \) and \( \mathbf{B}_2 \) are given by:
\[
\hat{\mathbf{B}}_1 = \mathbf{Y}_1 \hat{\boldsymbol{\Psi}}^{-1} \mathbf{X}_1^\top \left( \mathbf{X}_1 \hat{\boldsymbol{\Psi}}^{-1} \mathbf{X}_1^\top \right)^{-1},
\qquad
\hat{\mathbf{B}}_2 = \mathbf{Y}_2 \hat{\boldsymbol{\Psi}}^{-1} \mathbf{X}_2^\top \left( \mathbf{X}_2 \hat{\boldsymbol{\Psi}}^{-1} \mathbf{X}_2^\top \right)^{-1}.
\]

These expressions are structurally identical to the estimator derived in the fully dense case, with each block estimated independently based on its corresponding partition of the data.

\subsection[\texorpdfstring{$\boldsymbol{\Sigma}$}{Sigma}]{Estimation of $\boldsymbol{\Sigma}$}

To estimate \( \boldsymbol{\Sigma} \), we treat \( \mathbf{B} \) and \( \boldsymbol{\Psi} \) as fixed and define the residual matrix \( \mathbf{E} = \mathbf{Y} - \mathbf{B} \mathbf{X} \). Under this setup, the log-likelihood for \( \boldsymbol{\Sigma} \) simplifies to
\[
\ell(\boldsymbol{\Sigma}) =
- \frac{r}{2} \log |\boldsymbol{\Sigma}|
- \frac{1}{2} \mathrm{tr} \left( \boldsymbol{\Sigma}^{-1} \mathbf{E} \boldsymbol{\Psi}^{-1} \mathbf{E}^\top \right).
\]

Taking the derivative of \( \ell(\boldsymbol{\Sigma}) \) with respect to \( \boldsymbol{\Sigma} \) and applying standard results from matrix calculus yields
\[
\frac{\partial \ell(\boldsymbol{\Sigma})}{\partial \boldsymbol{\Sigma}} =
- \frac{r}{2} \boldsymbol{\Sigma}^{-1}
+ \frac{1}{2} \boldsymbol{\Sigma}^{-1} \mathbf{E} \boldsymbol{\Psi}^{-1} \mathbf{E}^\top \boldsymbol{\Sigma}^{-1}.
\]

Setting this derivative to zero and solving for \( \boldsymbol{\Sigma} \) leads to the ML estimator:
\[
\hat{\boldsymbol{\Sigma}} = \frac{1}{r} \mathbf{E} \hat{\boldsymbol{\Psi}}^{-1} \mathbf{E}^\top = \frac{1}{r} (\mathbf{Y} - \hat{\mathbf{B}} \mathbf{X}) \hat{\boldsymbol{\Psi}}^{-1} (\mathbf{Y} - \hat{\mathbf{B}} \mathbf{X})^\top.
\]

This estimator quantifies the residual covariance among the \( p \) variables after accounting for the covariate effects and the spatio-temporal dependence structure captured by \( \hat{\boldsymbol{\Psi}} \). Notably, \( \boldsymbol{\Sigma} \) remains unstructured, as it is designed to model covariance between variables, not the spatial or temporal dependence.

\subsection[\texorpdfstring{$\boldsymbol{\Psi}$}{Psi}]{Estimation of $\boldsymbol{\Psi}$}

The column covariance matrix \( \boldsymbol{\Psi} \in \mathbb{R}^{r \times r} \), which models spatio-temporal dependencies, is assumed to have the separable Kronecker structure:
\[
\boldsymbol{\Psi} = \boldsymbol{\Psi}_{\text{sp}} \otimes \boldsymbol{\Psi}_{\text{tp}},
\]
where \( \boldsymbol{\Psi}_{\text{sp}} \in \mathbb{R}^{L \times L} \) and \( \boldsymbol{\Psi}_{\text{tp}} \in \mathbb{R}^{T \times T} \) capture spatial and temporal dependence, respectively.

For the spatial component, \( \boldsymbol{\Psi}_{\text{sp}} \), we consider five parametric covariance functions: the Matérn, exponential, Gaussian, cubic, and spherical models. The temporal component, \( \boldsymbol{\Psi}_{\text{tp}} \), is modeled using a first-order autoregressive (AR(1)) structure.

The parameters are estimated by maximizing the matrix-normal log-likelihood. With the residual matrix \(\mathbf{E} = \mathbf{Y} - \mathbf{B} \mathbf{X}\), the relevant part of the log-likelihood is:
\[
\ell(\boldsymbol{\Psi}) \propto 
- \frac{p}{2} \log|\boldsymbol{\Psi}| 
- \frac{1}{2} \mathrm{tr} \!\left( \boldsymbol{\Sigma}^{-1} \mathbf{E} \boldsymbol{\Psi}^{-1} \mathbf{E}^\top \right).
\]

Substituting the separable Kronecker structure and parameterizing as:
\[
\boldsymbol{\Psi}_{\text{sp}} = \sigma_s^2 \, \mathbf{R}_{\text{sp}}(\boldsymbol{\theta}_s), \qquad
\boldsymbol{\Psi}_{\text{tp}} = \mathbf{R}_{\text{tp}}(\phi_t),
\]
we obtain the profiled estimators:

The spatial variance parameter $\sigma_s^2$ has closed-form solution:
\[
\widehat{\sigma}_s^{\,2}
= \frac{1}{pLT}\,
\operatorname{tr}\left(
\hat{\boldsymbol{\Sigma}}^{-1}\hat{\mathbf{E}}\,
\big(\hat{\mathbf{R}}_{\text{sp}}^{-1}\!\otimes\!\hat{\mathbf{R}}_{\text{tp}}^{-1}\big)
\hat{\mathbf{E}}^\top
\right).
\]

The spatial range parameter $\phi_s$ is estimated by solving:
\[
\frac{pT}{2}\,
\operatorname{tr}\left(
\mathbf{R}_{\text{sp}}^{-1}\,\frac{\partial \mathbf{R}_{\text{sp}}}{\partial \phi_s}
\right)
= \frac{1}{2\hat{\sigma}_s^2}\,
\operatorname{tr}\left(
\boldsymbol{\Sigma}^{-1}\mathbf{E}\,
\big(
\mathbf{R}_{\text{sp}}^{-1}\frac{\partial \mathbf{R}_{\text{sp}}}{\partial \phi_s}
\mathbf{R}_{\text{sp}}^{-1}\otimes \mathbf{R}_{\text{tp}}^{-1}
\big)\,
\mathbf{E}^\top
\right).
\]

For the Matérn model, the smoothness parameter $\nu$ is estimated by solving:
\[
\frac{pT}{2}\,
\operatorname{tr}\left(
\mathbf{R}_{\text{sp}}^{-1}\,\frac{\partial \mathbf{R}_{\text{sp}}}{\partial \nu}
\right)
= \frac{1}{2\hat{\sigma}_s^2}\,
\operatorname{tr}\left(
\boldsymbol{\Sigma}^{-1}\mathbf{E}\,
\big(
\mathbf{R}_{\text{sp}}^{-1}\frac{\partial \mathbf{R}_{\text{sp}}}{\partial \nu}
\mathbf{R}_{\text{sp}}^{-1}\otimes \mathbf{R}_{\text{tp}}^{-1}
\big)\,
\mathbf{E}^\top
\right).
\]

The temporal autocorrelation parameter $\rho$ is estimated by solving:
\[
\frac{pL}{2}
\operatorname{tr}\!\left(
\mathbf{R}_{\text{tp}}^{-1} \frac{\partial \mathbf{R}_{\text{tp}}}{\partial \rho}
\right)
= \frac{1}{2\hat{\sigma}_s^2}
\operatorname{tr}\!\left(
\boldsymbol{\Sigma}^{-1} \mathbf{E}
\left(
\mathbf{R}_{\text{sp}}^{-1} \otimes
\mathbf{R}_{\text{tp}}^{-1} \frac{\partial \mathbf{R}_{\text{tp}}}{\partial \rho} \mathbf{R}_{\text{tp}}^{-1}
\right)
\mathbf{E}^\top
\right).
\]

The specific forms of the derivatives required for the estimating equations are:

Exponential structure:
\[
\frac{\partial (\mathbf{R}_{\text{sp}})_{ij}}{\partial \phi_s} 
= \frac{h_{ij}}{\phi_s^2} \exp\left(-\frac{h_{ij}}{\phi_s}\right).
\]

Gaussian structure:
\[
\frac{\partial (\mathbf{R}_{\text{sp}})_{ij}}{\partial \phi_s} 
= \frac{2 h_{ij}^2}{\phi_s^{3}} \exp\left(-\frac{h_{ij}^2}{\phi_s^{2}}\right).
\]

Cubic structure:
\[
\frac{\partial (\mathbf{R}_{\text{sp}})_{ij}}{\partial \phi_s} = 
\begin{cases}
\frac{7 h_{ij}^2}{\phi_s^3} - \frac{35 h_{ij}^3}{4 \phi_s^4} + \frac{7 h_{ij}^5}{4 \phi_s^6}, & \text{for } 0 \leq h_{ij} \leq \phi_s, \\
0, & \text{for } h_{ij} > \phi_s.
\end{cases}
\]

Spherical structure:
\[
\frac{\partial (\mathbf{R}_{\text{sp}})_{ij}}{\partial \phi_s} = 
\begin{cases}
\frac{3 h_{ij}^2}{\phi_s^4} - \frac{9 h_{ij}^3}{4 \phi_s^5}, & \text{for } 0 \leq h_{ij} \leq \phi_s, \\
0, & \text{for } h_{ij} > \phi_s.
\end{cases}
\]

Matérn model:
\[
\frac{\partial (\mathbf{R}_{\text{sp}})_{ij}}{\partial \phi_s} = \frac{2^{1-\nu} h_{ij}}{\Gamma(\nu) \phi_s^2} \left(\frac{h_{ij}}{\phi_s}\right)^{\nu} K_{\nu-1}\!\left(\frac{h_{ij}}{\phi_s}\right),
\]
\[
\frac{\partial \mathbf{R}_{\text{sp}}(h_{ij})}{\partial \nu}
= \mathbf{R}_{\text{sp}}(h_{ij}) \left[
-\log(2) - \psi(\nu) + \log\!\left(\frac{h_{ij}}{\phi_s}\right)
+ \frac{\partial}{\partial \nu} \log K_\nu\!\left(\frac{h_{ij}}{\phi_s}\right)
\right].
\]

AR(1) temporal correlation:
\[
\frac{\partial (\mathbf{R}_{\text{tp}})_{tt'}}{\partial \rho} = 
|t - t'| \cdot \rho^{\,|t - t'| - 1}.
\]

These derivatives are essential components of the previously presented score equations, enabling the estimation of the spatial range parameter $\phi_s$ for all covariance structures, the smoothness parameter $\nu$ for the Matérn model, and the temporal autocorrelation parameter $\rho$.

The spatial variance parameter $\sigma_s^2$ has closed-form solution:
\[
\widehat{\sigma}_s^{\,2}
= \frac{1}{pLT}\,
\operatorname{tr}\left(
\hat{\boldsymbol{\Sigma}}^{-1}\hat{\mathbf{E}}\,
\big(\hat{\mathbf{R}}_{\text{sp}}^{-1}\!\otimes\!\hat{\mathbf{R}}_{\text{tp}}^{-1}\big)
\hat{\mathbf{E}}^\top
\right),
\]
where $\hat{\mathbf{E}} = \mathbf{Y} - \hat{\mathbf{B}}\mathbf{X}$ is the residual matrix based on current parameter estimates.

\section{Residual analysis and model checking}\label{resid}

For model validation, we examine the standardized residuals. Under the matrix-normal assumption $\mathbf{E} \sim \mathcal{N}_{p \times r}(\mathbf{0}, \boldsymbol{\Sigma}, \boldsymbol{\Psi})$, the vectorized residuals follow:
\[
\operatorname{vec}(\mathbf{E}) \sim \mathcal{N}_{pr}(\mathbf{0}, \boldsymbol{\Psi} \otimes \boldsymbol{\Sigma}).
\]

The standardized residuals are obtained by:
\[
\mathbf{E}^* = \hat{\boldsymbol{\Sigma}}^{-1/2} \hat{\mathbf{E}} \hat{\boldsymbol{\Psi}}^{-1/2},
\]
where $\mathbf{E}^*$ should approximately follow $\mathcal{N}_{p \times r}(\mathbf{0}, \mathbf{I}_p, \mathbf{I}_r)$ if the model is correctly specified.

Given that we typically have only one realization of the residual matrix, formal multivariate normality tests are not feasible. Instead, we employ graphical methods such as Q-Q plots of the standardized residuals and marginal checks via univariate normality tests on the elements of $\mathbf{E}^*$. Additionally, we can compute the inner product $\operatorname{vec}(\mathbf{E}^*)^\top \operatorname{vec}(\mathbf{E}^*)$ which, under the null hypothesis of correct model specification, should follow a chi-squared distribution with $pr$ degrees of freedom.

Local residual diagnostics for the matrix-variate model can be obtained from the residuals of the \(p\) responses at the corresponding space--time index. For \(j \in \{1, \dots, r\}\), let \(\mathbf{E}_{\cdot j} \in \mathbb{R}^{p}\) denote the \(j\)-th column of \(\mathbf{E}\).

Based on the matrix-normal distribution structure, it follows that:
\[
\mathbf{E}_{\cdot j} \sim \mathcal{N}_p\!\big(\mathbf{0},\, \psi_{jj} \, \boldsymbol{\Sigma}\big),
\qquad
\operatorname{Cov}\!\big(\mathbf{E}_{\cdot j}, \mathbf{E}_{\cdot k}\big) = \psi_{jk} \, \boldsymbol{\Sigma}.
\]
Here, \(\psi_{jj}\) is the \(j\)-th diagonal element of the column-covariance matrix \(\boldsymbol{\Psi} = (\psi_{jk})_{j,k=1}^r\). This element scales the row-covariance matrix \(\boldsymbol{\Sigma}\) for column \(j\), such that \(\mathrm{Var}(\mathbf{E}_{\cdot j}) = \psi_{jj} \, \boldsymbol{\Sigma}\).

Therefore, only \(\psi_{jj}\) appears as a scale factor in the marginal distribution of a single column, whereas the off-diagonal elements \(\psi_{jk}\) govern the dependence between different columns.

A natural local diagnostic for column \(j\) is the marginal Mahalanobis distance
\[
d_j^2
\;=\;
\frac{\mathbf{E}_{\cdot j}^{\!\top}\,\widehat{\boldsymbol{\Sigma}}^{-1}\,\mathbf{E}_{\cdot j}}
{\widehat{\psi}_{jj}}
\;\approx\; \chi^2_{p},
\]
which measures the joint deviation of the \(p\) responses at that space--time index. Large values (e.g., above the \(97.5\%\) quantile of \(\chi^2_p\)) flag potential local outliers.

Complementary to the column-wise view, row-wise and cell-level diagnostics provide insights at different granularities:
\[
r_i^2
\;=\;
\frac{\mathbf{E}_{i\cdot}\,\widehat{\boldsymbol{\Psi}}^{-1}\,\mathbf{E}_{i\cdot}^{\!\top}}
{\widehat{\sigma}_{ii}}
\;\approx\; \chi^2_{r},
\qquad
z_{ij}
\;=\;
\frac{e_{ij}}{\sqrt{\widehat{\sigma}_{ii}\,\widehat{\psi}_{jj}}}
\;\approx\; \mathcal{N}(0,1).
\]

Specifically, the row-wise statistic \(r_i^2\) assesses the overall unusualness of all responses for a given experimental unit or location \(i\) across all \(r\) time points (or conditions). A large \(r_i^2\) value indicates that the entire profile of row \(i\) is anomalous, potentially flagging a consistently aberrant unit throughout the experiment.    The cell-level residual \(z_{ij}\) isolates the standardized deviation of a single response for unit \(i\) at a specific time/condition \(j\). This is the most granular diagnostic, useful for pinpointing the exact location of an outlier within the data matrix once a suspicious row or column has been identified.

Thus, \(d_j^2\) detects anomalous time points/conditions, \(r_i^2\) detects anomalous experimental units, and \(z_{ij}\) pinpoints individual anomalous measurements. Note that \(\|\mathbf{E}^*_{\cdot j}\|^2 = d_j^2\) only when \(\widehat{\boldsymbol{\Psi}}\) is diagonal; otherwise, right-whitening by \(\widehat{\boldsymbol{\Psi}}^{-1/2}\) mixes columns and is not appropriate for per-column diagnostics.

\section{Simulation studies}\label{sec:simulation} 

This section presents a simulation study designed to evaluate the performance of the proposed matrix-variate regression model in recovering its parameters under various spatio-temporal configurations. We focus on assessing the accuracy of the estimates for the regression coefficients, the row covariance matrix, and the parameters of the separable covariance structure.

Data were generated from the model
\[
\mathbf{Y} = \mathbf{B}\mathbf{X} + \mathbf{E}, 
\qquad 
\mathbf{E} \sim \mathcal{N}_{p \times r}\!\left(\mathbf{0}, \boldsymbol{\Sigma}, \boldsymbol{\Psi}\right),
\]
with \( p=3 \), \( q=3 \), \( r = L \times T \), and the separable covariance structure \( \boldsymbol{\Psi} = \boldsymbol{\Psi}_{\text{sp}} \otimes \boldsymbol{\Psi}_{\text{tp}} \). The covariate matrix \( \mathbf{X} \in \mathbb{R}^{q \times r} \) was populated with independent standard normal entries.

The matrix of regression coefficients was fixed at
\[
\mathbf{B} =
\begin{bmatrix}
1.00 & 1.40 & 2.00 \\ 
1.00 & 1.20 & 1.00 \\ 
2.00 & 1.00 & 1.20 
\end{bmatrix},
\]
and the row covariance matrix \( \boldsymbol{\Sigma} \in \mathbb{R}^{p \times p} \) followed an AR(1) structure with autocorrelation 0.4:
\[
\boldsymbol{\Sigma} =
\begin{bmatrix}
1.00 & 0.40 & 0.16 \\
0.40 & 1.00 & 0.40 \\
0.16 & 0.40 & 1.00
\end{bmatrix}.
\]

The spatial component \( \boldsymbol{\Psi}_{\text{sp}} \in \mathbb{R}^{L \times L} \) employed the five correlation structures detailed in Section~\ref{sec:reg_model}. All spatial models shared the common parameterization \(\sigma_s^2 = 1.1\) and \(\phi_s = 1.2\), enabling a direct comparison of their performance. For the Mat\'ern covariance structure, the smoothness parameter was set to \(\nu = 1.5\). The temporal component \( \boldsymbol{\Psi}_{\text{tp}} \in \mathbb{R}^{T \times T} \) used an AR(1) structure with autocorrelation \(\rho = 0.7\) and unit variance.

To assess performance under different data structures, we defined two main settings. The first fixed the temporal dimension (\(T=12\)) and varied the number of spatial locations \(L \in \{5, 10, 20\}\), while the second fixed the spatial dimension (\(L=12\)) and varied the time points \(T \in \{6, 12, 24\}\). For each value of \(L\), the spatial locations were randomly sampled from a regular grid over \([1, 10] \times [1, 10]\).

For each scenario, we performed 100 Monte Carlo replications. Performance was evaluated using the Frobenius norm to measure the estimation error of the matrix parameters $\mathbf{B}$ and $\boldsymbol{\Sigma}$, and the mean squared error (MSE) to assess the accuracy of the scalar parameters $\sigma_s^2$, $\phi_s$, $\rho$, and $\nu$.

\begin{figure}[H]
\centering

\begin{subfigure}[t]{0.24\textwidth}
    \centering
    \makebox[0pt]{\textbf{$\mathbf{B}$ ($T=12$)}}\\[0.5ex]
    \includegraphics[width=\textwidth]{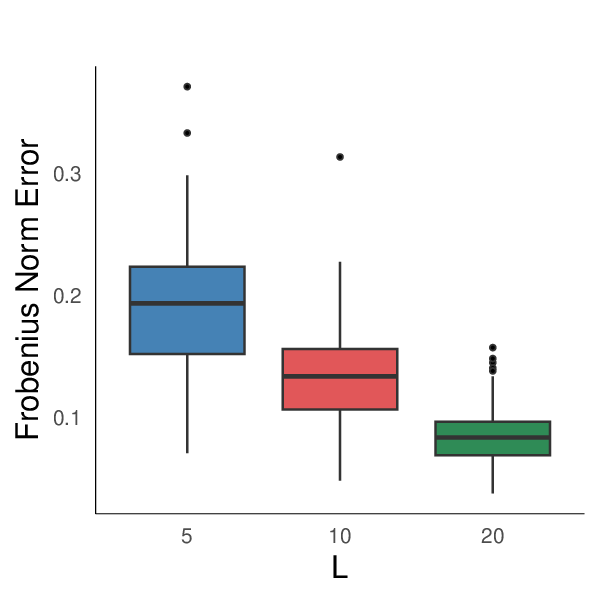}
    {\textbf{(a)}}
\end{subfigure}\hfill
\begin{subfigure}[t]{0.24\textwidth}
    \centering
    \makebox[0pt]{\textbf{ $\boldsymbol{\Sigma}$ ($T=12$)}}\\[0.5ex]
    \includegraphics[width=\textwidth]{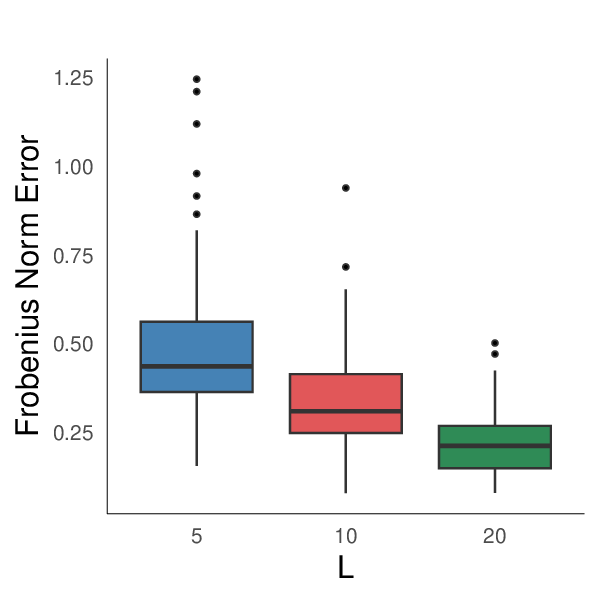}
    {\textbf{(b)}}
\end{subfigure}\hfill
\begin{subfigure}[t]{0.24\textwidth}
    \centering
    \makebox[0pt]{\textbf{ $\mathbf{B}$ ($L=12$)}}\\[0.5ex]
    \includegraphics[width=\textwidth]{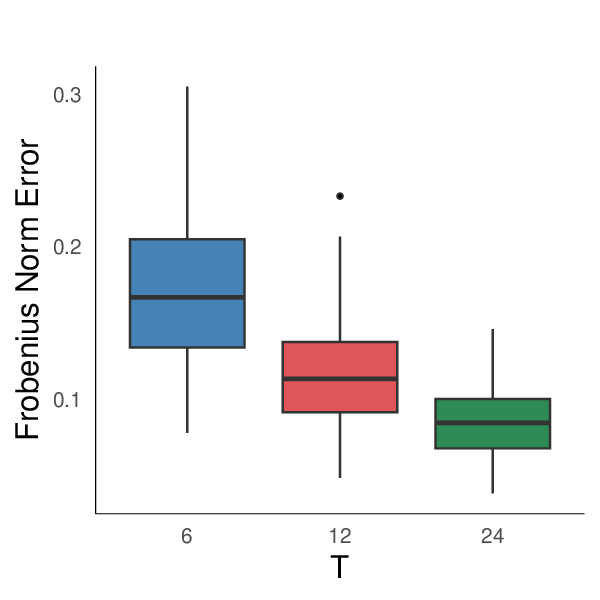}
    {\textbf{(c)}}
\end{subfigure}\hfill
\begin{subfigure}[t]{0.24\textwidth}
    \centering
    \makebox[0pt]{\textbf{ $\boldsymbol{\Sigma}$ ( $L=12$)}}\\[0.5ex]
    \includegraphics[width=\textwidth]{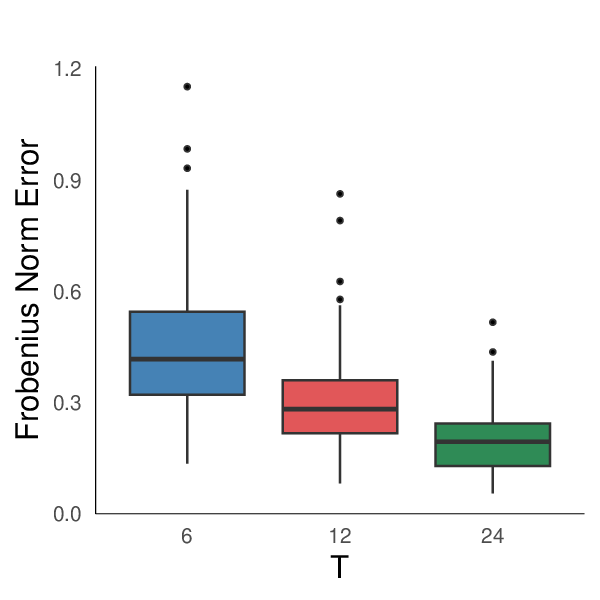}
    {\textbf{(d)}}
    
\end{subfigure}

\caption{Frobenius norm errors for the coefficient matrix $\mathbf{B}$ and the covariance matrix $\boldsymbol{\Sigma}$ under the Exponential covariance: fixed $T$ (a–b) and fixed $L$ (c–d).}
\label{fig:exp_frobenius_all}
\end{figure}

The results for the Exponential covariance structure with fixed \(T = 12\) illustrate the general behavior observed across all experimental designs and covariance settings (Gaussian, Matérn, cubic, and spherical), as summarized in Figure~\ref{fig:exp_frobenius_all}. Under this configuration, the estimated coefficient matrices \(\widehat{\mathbf B}\) closely match the true values \(\mathbf B_0\) for all \(L \in \{5, 10, 20\}\), with Frobenius norm errors ranging from \(0.014\) to \(0.017\), indicating no systematic bias. Likewise, the estimated row covariance \(\widehat{\boldsymbol{\Sigma}}\) accurately reproduces the target structure across all scenarios, showing only a mild positive bias along the diagonal and off-diagonal discrepancies within \(\pm 0.02\). A similar pattern is observed when \(L\) is fixed and \(T\) varies, consistently replicated across the Gaussian, Matérn, cubic, and spherical covariance structures, confirming the stability and reliability of parameter estimation under all configurations.

Figure~\ref{fig:exp_frobenius_all}  compiles all boxplots: panels (a–b) fix $T=12$ ($n=100$) and show Frobenius‐norm errors for $\widehat{\mathbf B}$ and $\widehat{\boldsymbol{\Sigma}}$ declining as $L$ increases from $5$ to $20$, with the largest gains from $L=5$ to $L=10$ and $\widehat{\boldsymbol{\Sigma}}$ remaining harder to estimate but following the same monotonic pattern; panels (c–d) fix $L=12$ and display the analogous behavior as $T$ grows from $6$ to $24$, with the biggest drop between $T=6$ and $T=12$, confirming that accuracy improves with either more locations or more time points.

\begin{figure}[H]
\centering

\begin{subfigure}[t]{0.24\textwidth}
    \centering
    \makebox[0pt]{\textbf{$\mathbf{B}$ ($T=12$)}}\\[0.5ex]
    \includegraphics[width=\textwidth]{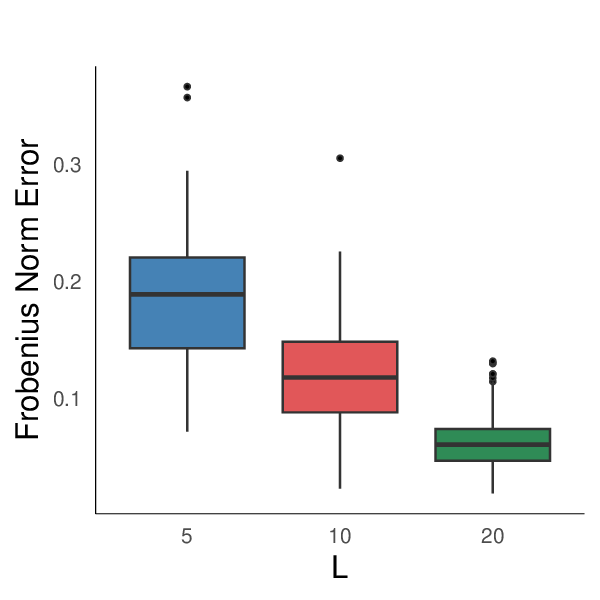}
    {\textbf{(a)}}
\end{subfigure}\hfill
\begin{subfigure}[t]{0.24\textwidth}
    \centering
    \makebox[0pt]{\textbf{ $\boldsymbol{\Sigma}$ ($T=12$)}}\\[0.5ex]
    \includegraphics[width=\textwidth]{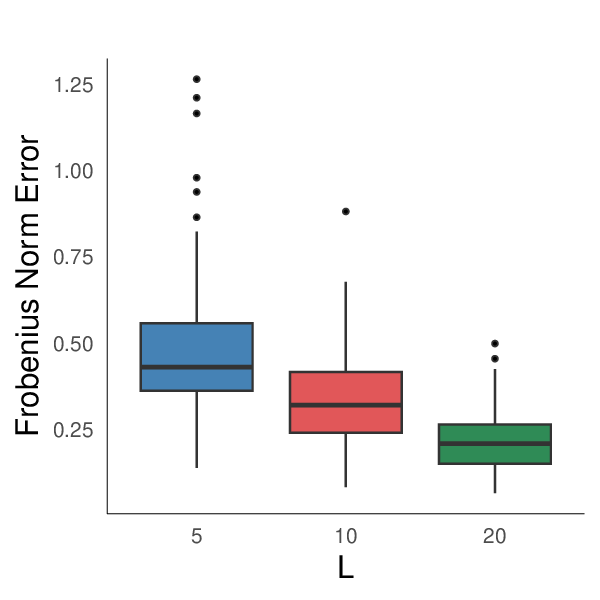}
    {\textbf{(b)}}
\end{subfigure}\hfill
\begin{subfigure}[t]{0.24\textwidth}
    \centering
    \makebox[0pt]{\textbf{ $\mathbf{B}$ ($L=12$)}}\\[0.5ex]
    \includegraphics[width=\textwidth]{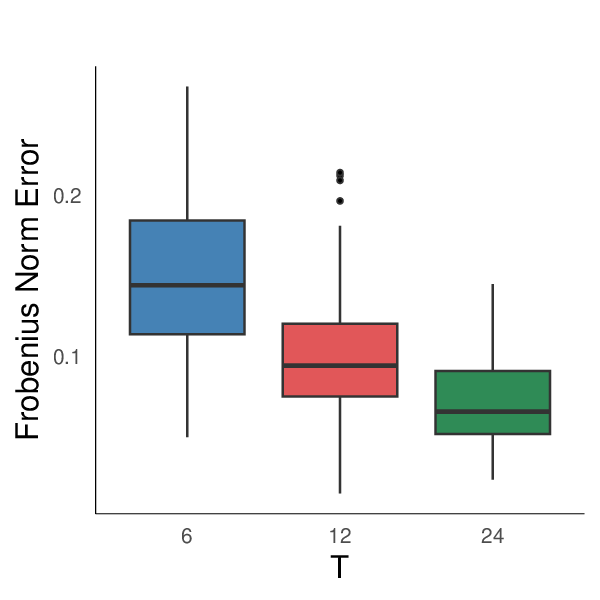}
    {\textbf{(c)}}
\end{subfigure}\hfill
\begin{subfigure}[t]{0.24\textwidth}
    \centering
    \makebox[0pt]{\textbf{ $\boldsymbol{\Sigma}$ ( $L=12$)}}\\[0.5ex]
    \includegraphics[width=\textwidth]{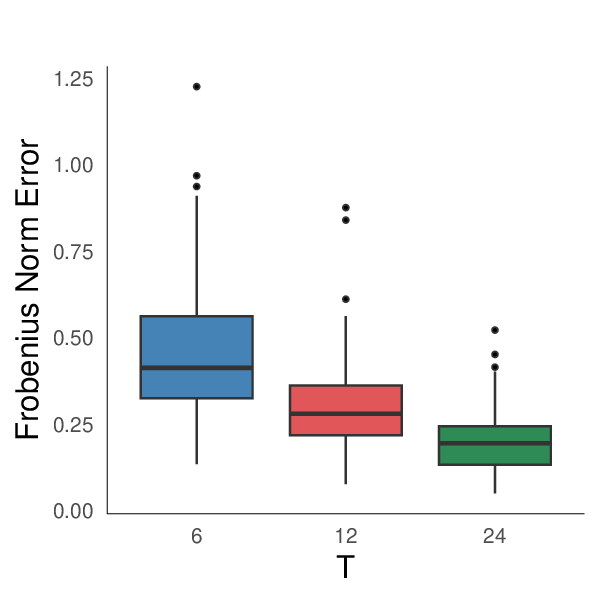}
    {\textbf{(d)}}
\end{subfigure}

\caption{Frobenius norm errors for the coefficient matrix $\mathbf{B}$ and the covariance matrix $\boldsymbol{\Sigma}$ under the Gaussian covariance: fixed $T$ (a–b) and fixed $L$ (c–d).}
\label{fig:gaus_frobenius_all}
\end{figure}

The Gaussian counterpart in Figure~\ref{fig:gaus_frobenius_all} exhibits the same qualitative pattern under both fixed $T$ and fixed $L$, and identical behavior (figures omitted) is observed for the Matérn, cubic, and spherical structures, indicating robust parameter recovery across covariance specifications and designs.

\begin{figure}[H]
\centering

\begin{subfigure}[b]{0.240\textwidth}
  \centering
  \makebox[0pt]{\textbf{$\sigma^2$ ($T=12$)}}\\[0.5ex]

  \includegraphics[width=\textwidth]{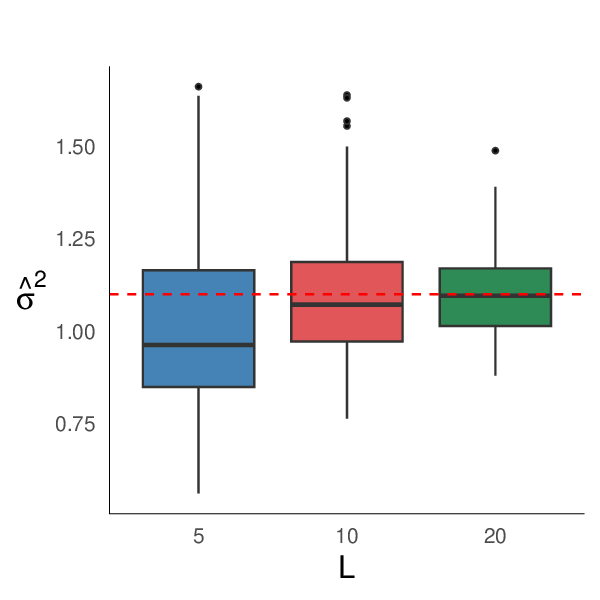}
  {\textbf{(a)}}
  \label{fig:sigma2_T_exp}
\end{subfigure}\hfill
\begin{subfigure}[b]{0.240\textwidth}
  \centering
  \makebox[0pt]{\textbf{$\phi$ ($T=12$)}}\\[0.5ex]
  \includegraphics[width=\textwidth]{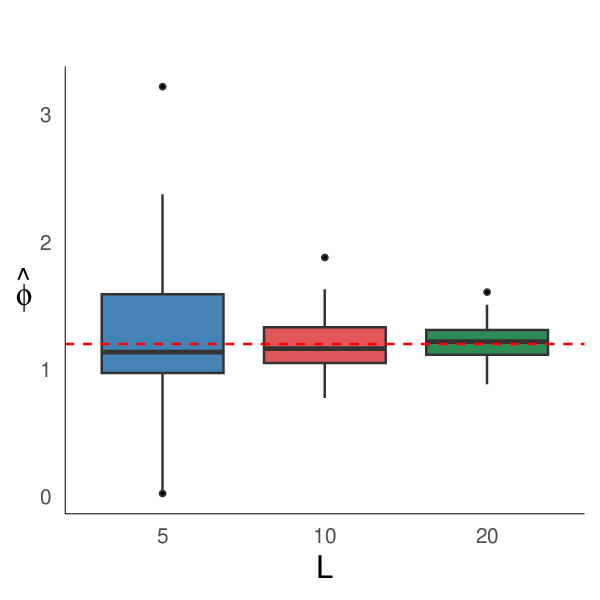}
   {\textbf{(b)}}
  \label{fig:phi_T_exp}
\end{subfigure}\hfill
\begin{subfigure}[b]{0.240\textwidth}
  \centering
  \makebox[0pt]{\textbf{$\rho$ ($T=12$)}}\\[0.5ex]
  \includegraphics[width=\textwidth]{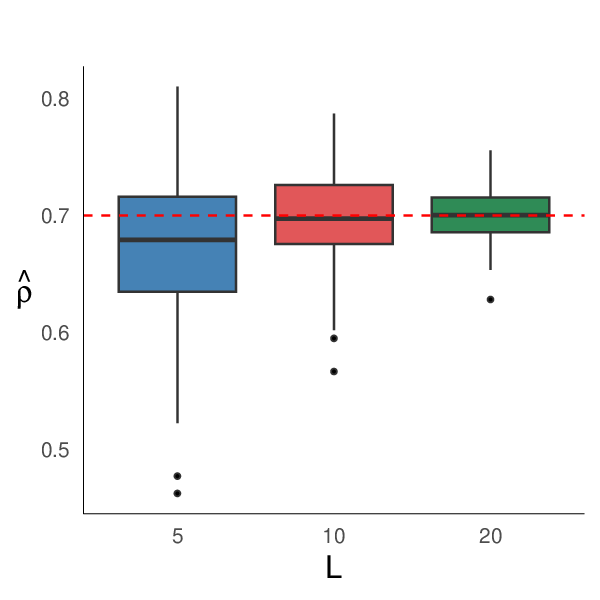}
  {\textbf{(c)}}
  \label{fig:rho_T_exp}
\end{subfigure}

\vspace{5pt}

\begin{subfigure}[b]{0.240\textwidth}
  \centering
  \makebox[0pt]{\textbf{$\sigma^2$ ($L=12$)}}\\[0.5ex]
  \includegraphics[width=\textwidth]{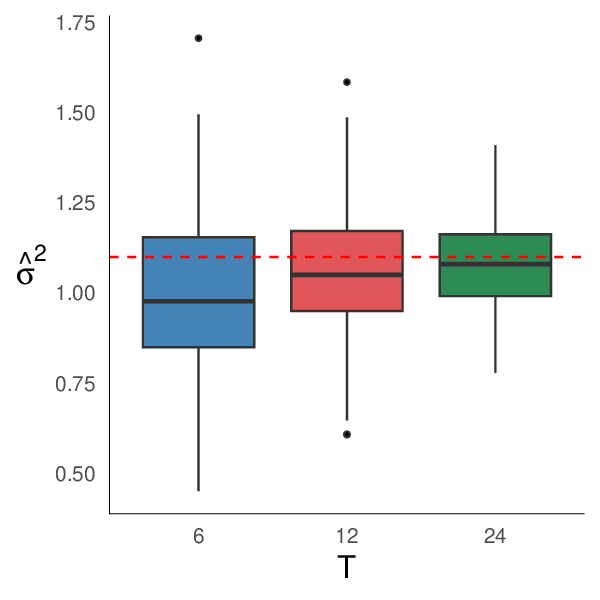}
  {\textbf{(d)}}
  \label{fig:sigma2_L_exp}
\end{subfigure}\hfill
\begin{subfigure}[b]{0.240\textwidth}
  \centering
  \makebox[0pt]{\textbf{$\phi$ ($L=12$)}}\\[0.5ex]
  \includegraphics[width=\textwidth]{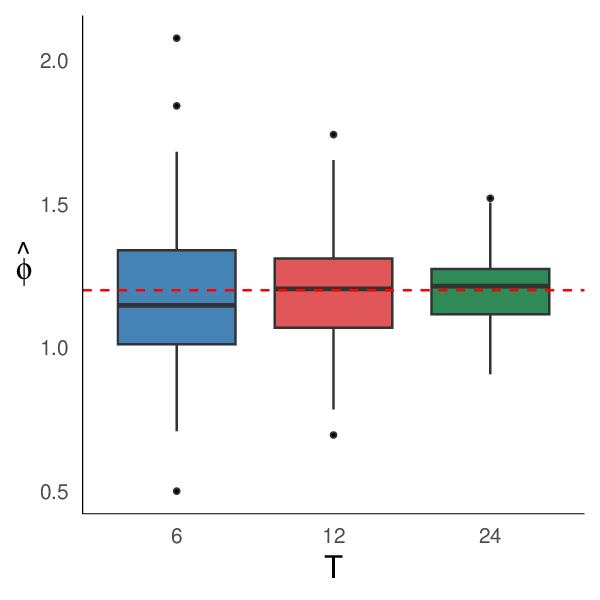}
  {\textbf{(e)}}
  \label{fig:phi_L_exp}
\end{subfigure}\hfill
\begin{subfigure}[b]{0.240\textwidth}
  \centering
  \makebox[0pt]{\textbf{$\rho$ ($L=12$)}}\\[0.5ex]
  \includegraphics[width=\textwidth]{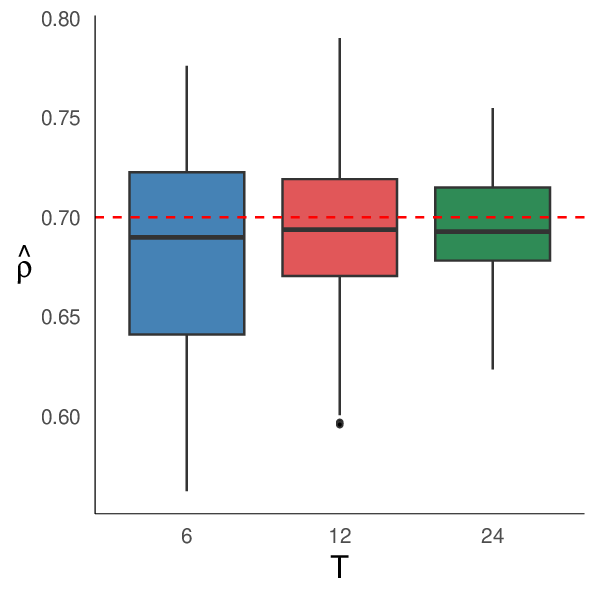}
 {\textbf{(f)}x}
  \label{fig:rho_L_exp}
\end{subfigure}

\caption{Estimated values for the scalar parameters under the exponential covariance: top row (a–c) fixes $L$, and bottom row (d–f) fixes $T$. The red dashed line indicates the true values $\sigma^2=1.1$, $\phi_s=1.2$, and $\rho=0.7$.}
\label{fig:exp_scalar_params_combined}
\end{figure}

\begin{figure}[H]
\centering

\begin{subfigure}[b]{0.240\textwidth}
  \centering
  \makebox[0pt]{\textbf{$\sigma^2$ ($T=12$)}}\\[0.5ex]

  \includegraphics[width=\textwidth]{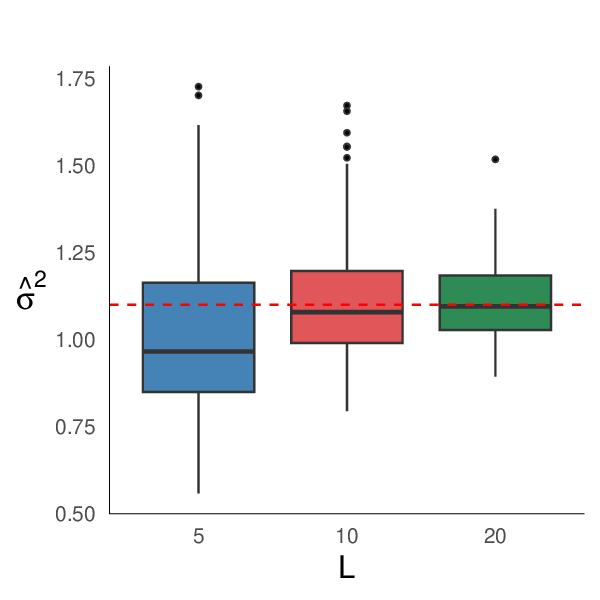}
  {\textbf{(a)}}
  \label{fig:sigma2_T_gaussian}
\end{subfigure}\hfill
\begin{subfigure}[b]{0.240\textwidth}
  \centering
  \makebox[0pt]{\textbf{$\phi$ ($T=12$)}}\\[0.5ex]
  \includegraphics[width=\textwidth]{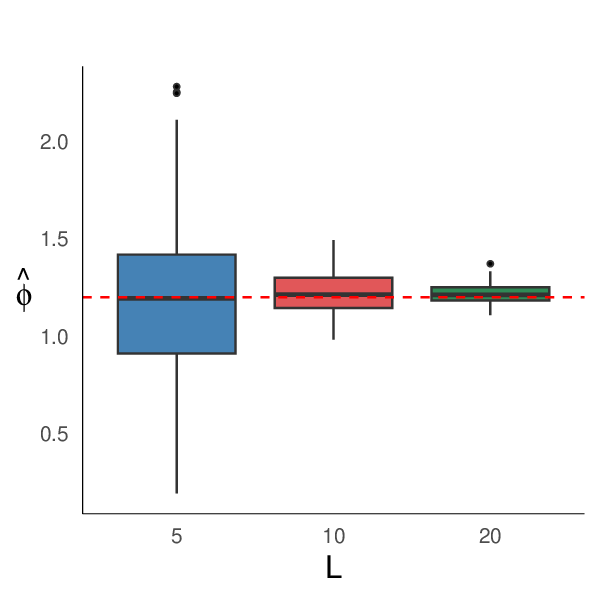}
   {\textbf{(b)}}
  \label{fig:phi_T_gaussian}
\end{subfigure}\hfill
\begin{subfigure}[b]{0.240\textwidth}
  \centering
  \makebox[0pt]{\textbf{$\rho$ ($T=12$)}}\\[0.5ex]
  \includegraphics[width=\textwidth]{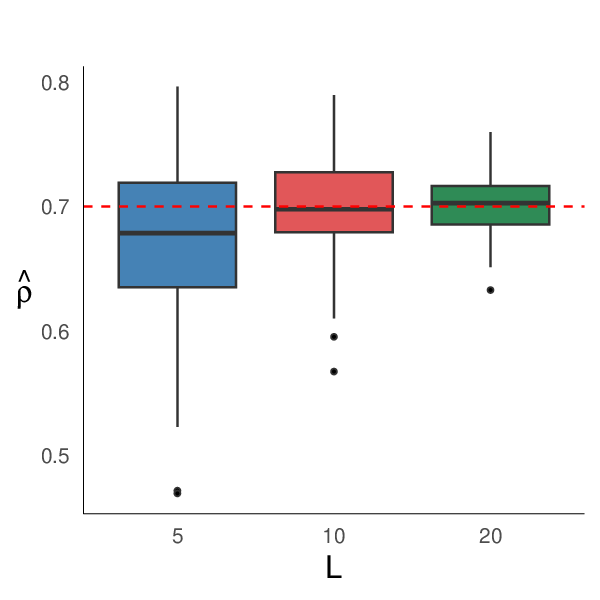}
  {\textbf{(c)}}
  \label{fig:rho_T_gaussian}
\end{subfigure}

\vspace{5pt}

\begin{subfigure}[b]{0.240\textwidth}
  \centering
  \makebox[0pt]{\textbf{$\sigma^2$ ($L=12$)}}\\[0.5ex]
  \includegraphics[width=\textwidth]{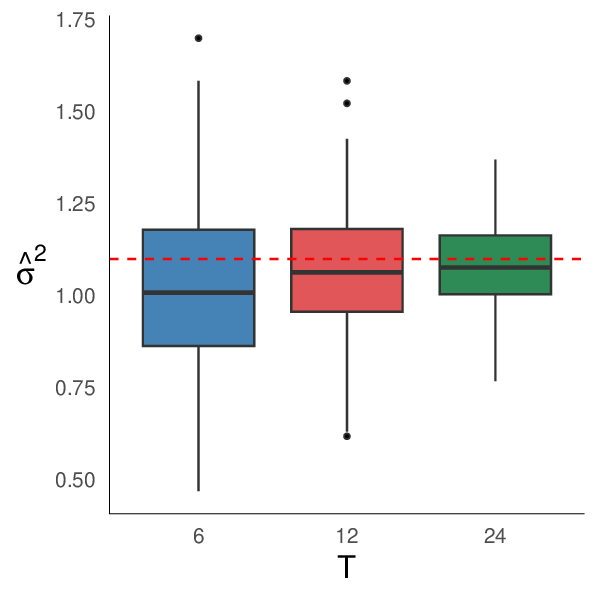}
  {\textbf{(d)}}
  \label{fig:sigma2_L_gaussian}
\end{subfigure}\hfill
\begin{subfigure}[b]{0.240\textwidth}
  \centering
  \makebox[0pt]{\textbf{$\phi$ ($L=12$)}}\\[0.5ex]
  \includegraphics[width=\textwidth]{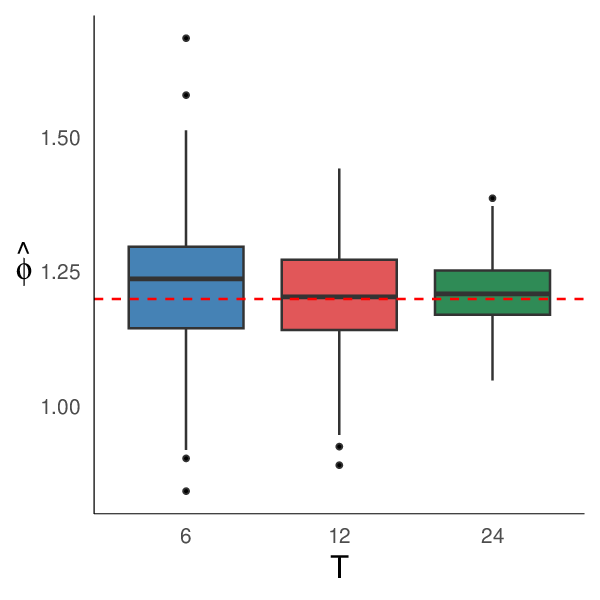}
  {\textbf{(e)}}
  \label{fig:phi_L_gaussian}
\end{subfigure}\hfill
\begin{subfigure}[b]{0.240\textwidth}
  \centering
  \makebox[0pt]{\textbf{$\rho$ ($L=12$)}}\\[0.5ex]
  \includegraphics[width=\textwidth]{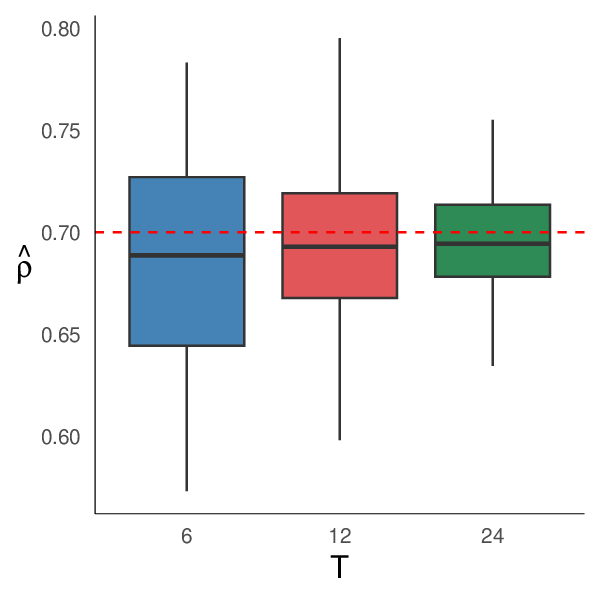}
 {\textbf{(f)}}
  \label{fig:rho_L_gaussian}
\end{subfigure}

\caption{Estimated values for the scalar parameters under the Gaussian covariance: top row (a–c) fixes $L$, and bottom row (d–f) fixes $T$. The red dashed line indicates the true values $\sigma^2=1.1$, $\phi_s=1.2$, and $\rho=0.7$.}
\label{fig:gau_scalar_params_combined}
\end{figure}

Figure~\ref{fig:exp_scalar_params_combined} (Exponential) shows accurate recovery of the scalar parameters $\sigma_s^2$, $\phi_s$, and $\rho$: with $T$ fixed, increasing $L$ (top row, a–c) reduces dispersion and brings medians to the red dashed truths, and with $L$ fixed, increasing $T$ (bottom row, d–f) yields the same monotonic improvement (largest from 6$\to$12). The Gaussian counterpart in Figure~\ref{fig:gau_scalar_params_combined} exhibits the same qualitative pattern under both designs; analogous behavior (figures omitted) holds for the Matérn, cubic, and spherical structures, indicating robust parameter recovery across covariance specifications.

\begin{figure}[H]
\centering

\begin{subfigure}[t]{0.40\textwidth}
    \centering
    \makebox[0pt]{\textbf{$\mathbf{B}$ ($T=12$)}}\\[0.5ex]
    \includegraphics[width=\textwidth]{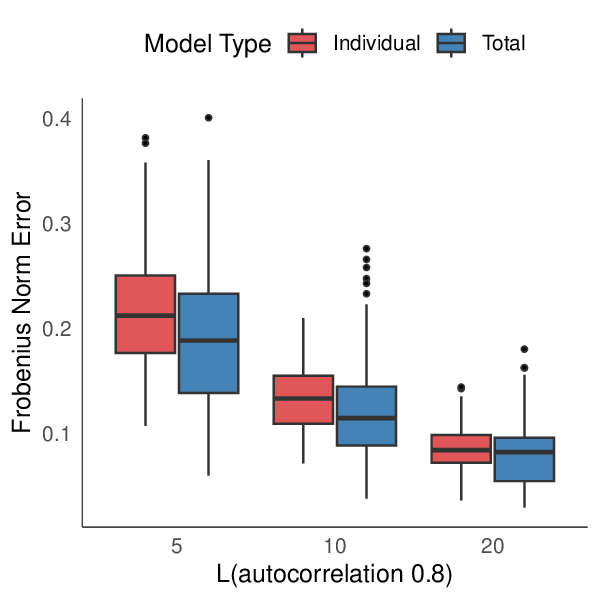}
    \caption{}
    \label{fig:exp_beta_T_v2}
\end{subfigure}\hfill
\begin{subfigure}[t]{0.40\textwidth}
    \centering
    \makebox[0pt]{\textbf{$\mathbf{B}$ ($L=12$)}}\\[0.5ex]
    \includegraphics[width=\textwidth]{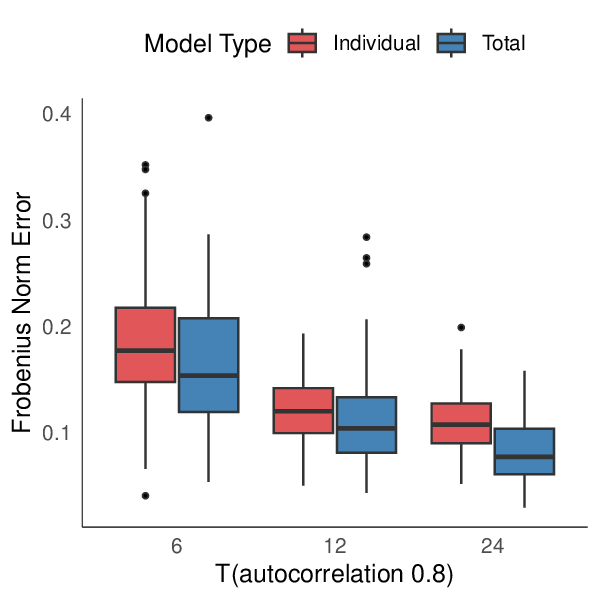}
    \caption{}
    \label{fig:exp_beta_T_v3}
\end{subfigure}

\caption{Frobenius norm errors $\|\widehat{\mathbf{B}}-\mathbf{B}\|_F$ for the estimated coefficient matrix. 
Blue boxplots refer to the joint (matrix–variate) regression where all responses are modeled together, 
and red boxplots to separate regressions fitted for each response. }

\label{fig:exp_frobenius_combined}
\end{figure}

Figure~\ref{fig:exp_frobenius_combined} presents the Frobenius norm errors between the estimated and true coefficient matrices $\widehat{\mathbf{B}}$ and $\mathbf{B}$, respectively, under two estimation strategies. The blue boxplots correspond to the matrix–variate regression model, in which all responses are modeled jointly (Total) and the coefficient matrix $\mathbf{B}$ is estimated while accounting for the covariance structure across both rows (response variables) and columns (spatial or temporal units).
The red boxplots, in contrast, correspond to the approach that fits independent regression models for each response variable in each scenario, after which the full coefficient matrix $\widehat{\mathbf{B}}$ is reconstructed by stacking the individually estimated coefficients. In both cases, the Frobenius norm error $\|\widehat{\mathbf{B}}-\mathbf{B}\|_F$ measures the overall deviation between the estimated and true coefficient matrices.

The left panel of Figure~\ref{fig:exp_frobenius_combined} shows the analysis with a fixed number of time points ($T=12$) and varying spatial locations ($L$). The joint (blue) model attains smaller median errors than the individual (red) fits, indicating that exploiting cross–response dependence enhances estimation efficiency, though at the cost of slightly higher variability.

The right panel of Figure~\ref{fig:exp_frobenius_combined} displays the complementary case with fixed $L=12$ and varying $T$. Increasing $T$ reduces the Frobenius errors for both approaches, showing the benefit of additional temporal information. Again, the joint model achieves lower median errors while presenting somewhat wider interquartile ranges, consistent with the bias–variance trade-off in shared estimation.

Overall, these results confirm that jointly modeling responses under the matrix–variate regression framework improves the estimation accuracy of $\mathbf{B}$ by properly accounting for the dependence structure among responses.

\section{Application}\label{sec:application}.

Table~\ref{tab:YXdata} in the Appendix summarizes key agricultural and livestock production metrics ($\mathbf{Y}{\text{year}}$) and associated covariates ($\mathbf{X}{\text{year}}$) for six municipalities in Minas Gerais, Brazil (Gonzaga, Guanhães, Patos de Minas, Prata, Uberaba, and Uberlândia) over the period from 2002 to 2021. The data are drawn from the Municipal Agricultural Production Survey (PAM) by IBGE (\cite{ibge2021pam}), which provides extensive agricultural data for all Brazilian municipalities. The complete dataset consists of 20 pairs of matrices, one for each year, where each pair represents the response and covariate information for a specific time point.

Each pair is composed of a response matrix $\mathbf{Y}_{\text{year}}$, containing agricultural indicators of interest, and a covariate matrix $\mathbf{X}_{\text{year}}$, containing associated socioeconomic or environmental variables. This structure allows for a direct application of the matrix-variate regression model under spatio-temporal dependence.

The original data were collected in absolute units (tons, heads, hectares, and BRL). To enhance readability and ensure consistent scaling across variables, the values were divided by appropriate powers of ten; all values in $\mathbf{Y}$ and the first two columns of $\mathbf{X}$ (temporary and permanent crops) were divided by $10^6$, while GDP values were divided by $10^9$.

\begin{itemize}
\item $\mathbf{Y}_{\text{year}}$: A $6 \times 3$ matrix containing:
\begin{enumerate}
\item Sugarcane production (in millions of tons),
\item Forestry production (in millions of tons),
\item Cattle herd size (in millions of heads).
\end{enumerate}

\item $\mathbf{X}_{\text{year}}$: A $6 \times 3$ matrix containing:
\begin{enumerate}
    \item Harvested area of temporary crops (in millions of hectares),
    \item Harvested area of permanent crops (in millions of hectares),
    \item Gross Domestic Product (GDP), in billions of BRL.
\end{enumerate}
\end{itemize}

\begin{figure}[H]
    \centering
    \includegraphics[width=0.9\textwidth]{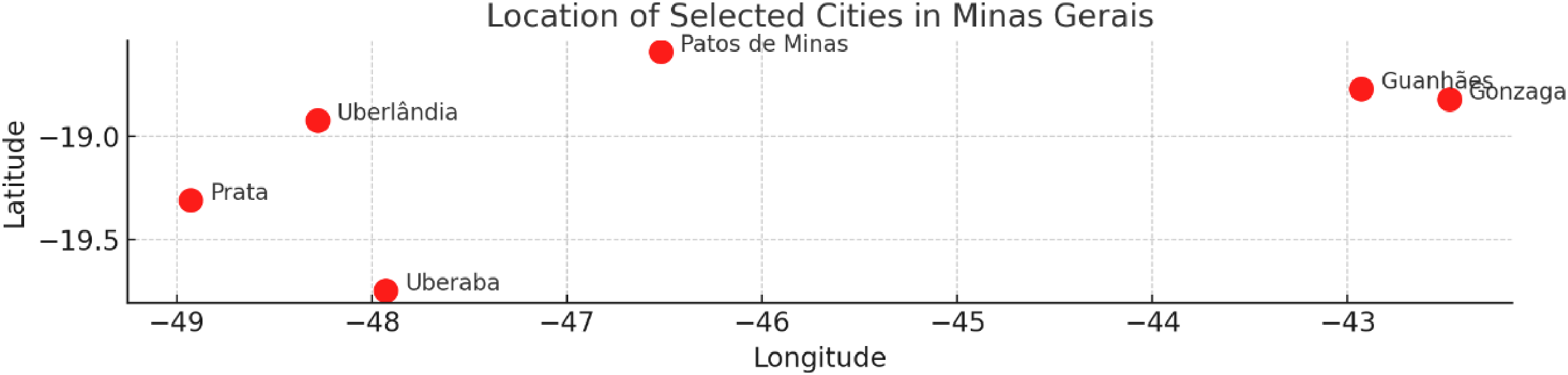}
    \caption{Brazilian agricultural data. Location of selected cities in Minas Gerais.}
    \label{fig:map_mg_cities}
\end{figure}

Figure~\ref{fig:map_mg_cities} displays the geographical locations of the six municipalities analyzed in this study within the state of Minas Gerais, Brazil. The red dots indicate the coordinates of each city, revealing their spatial distribution across the state's regions. This spatial arrangement underscores the dataset's geographic diversity, which is relevant for exploring potential spatial effects in the agricultural and economic indicators under study.

\begin{figure}[!ht]
  \centering
  \includegraphics[width=\linewidth]{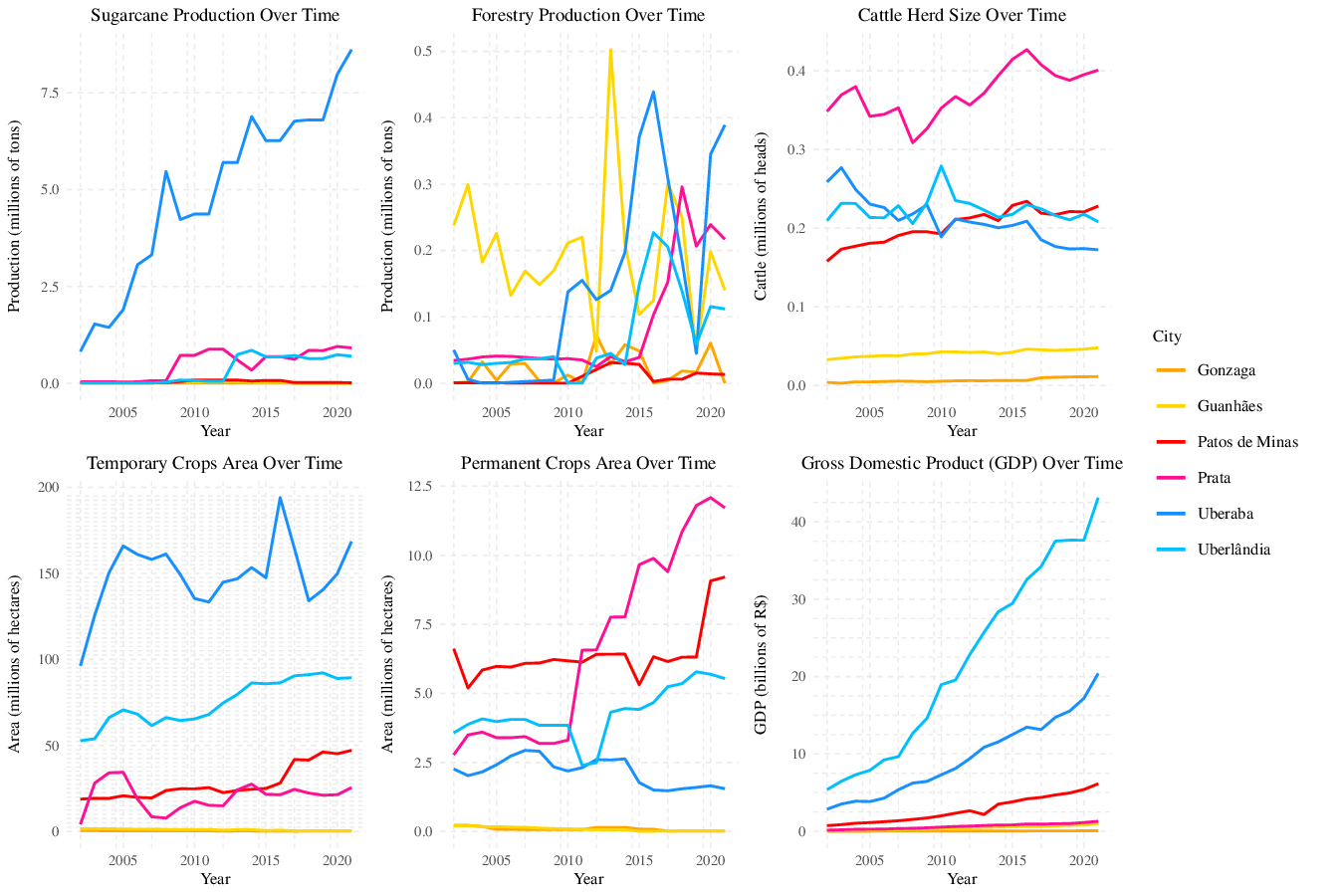}
  \caption{Brazilian agricultural data. Agricultural and economic indicators over time by location.}
  \label{Ind_over_time}
\end{figure}

The plots presented in Figure~\ref{Ind_over_time} provide a clear view of the temporal evolution of agricultural and economic indicators across different cities in Minas Gerais over a 20-year period. A marked heterogeneity is observed across locations: while some cities show strong growth in variables such as sugarcane production, cattle herd size, and GDP, others remain relatively stable. 

This spatial variation in growth patterns reflects differences in investment, infrastructure, agricultural vocations, and local economic dynamics. Furthermore, the explanatory variables (covariates) also exhibit distinct trajectories over time, which may directly influence the modeling of the response variables.

These observations motivate the use of a matrix-variate regression model with responses indexed by spatial locations and time points. This modeling framework captures the dependence structure through a separable covariance specification: one covariance matrix accounts for dependencies among response variables, while the other incorporates both spatial and temporal effects via a Kronecker product.

After formulating the matrix regression model, the analysis proceeded to model fitting across five spatial correlation structures. After fitting all five models, the Bayesian Information Criterion (BIC) was used to compare and select the best model. The BIC values for each candidate structure are presented in Table~\ref{tab:bic_compact}.

\begin{table}[H]
\centering
\caption{Brazilian agricultural data. BIC criteria for selecting the model under different correlation structures.}
\label{tab:bic_compact}
\small
\begin{tabular}{@{} l r @{}}
\toprule
Correlation & BIC \\
\midrule
Exponential & \textbf{-690.905} \\
Gaussian    & -690.665 \\
Cubic       & -690.718 \\
Spherical   & -690.360 \\
Matérn      & -690.721 \\
\bottomrule
\end{tabular}
\end{table}

As shown in Table~\ref{tab:bic_compact}, the model with the Exponential correlation structure achieved the lowest BIC, indicating it is the most appropriate among the candidates for describing the spatial dependence in this dataset. This finding suggests that the Exponential structure provides the optimal balance between goodness-of-fit and model complexity for our data. For this reason, all subsequent analyses and inferences presented in this work will be based exclusively on the exponential correlation structure model.

Our modeling strategy proceeded in two stages: an initial exploratory phase followed by a formal model selection. We fitted three competing models: Model 1 (dense $\mathbf{B}$), Model 2 ($\beta_{23} = 0$), and Model 3 ($\beta_{22} = \beta_{23} = 0$). The process began by fitting the full, unconstrained model (Model 1). An analysis of its parameter estimates suggested that coefficients $\beta_{23}$ and $\beta_{22}$ were negligible. While this initial assessment is inherently data-driven, it generated hypotheses to achieve model parsimony. Based on this exploratory observation, we proposed the two more parsimonious, nested models (Models 2 and 3). The final model was then selected from among these three candidates using the Bayesian Information Criterion (BIC). This criterion objectively balances model fit and complexity, helping to guard against overfitting. Therefore, while the candidate set was informed by the data, the final selection was guided by a rigorous, penalized-likelihood principle. Parameter estimates for Models 1 and 2 are presented in Table~\ref{parameter_est}.

The BIC values were $-691.091$ (Model 1), $-696.824$ (Model 2), and $-694.922$ (Model 3). Both constrained models improved upon Model 1, with Model 2 achieving the lowest BIC. This suggests that setting $\beta_{23} = 0$ provides the best balance, while the additional constraint in Model 3 is overly restrictive. With Model 2 established as the optimal model by the BIC, all subsequent analyses, including the interpretation of parameter estimates and diagnostic checks of the residuals, are specific to this chosen specification.

As discussed in Section~\ref{resid}, formal multivariate tests are not viable with a single residual matrix realization. For a correctly specified matrix-normal model, the Mahalanobis radius/chi-square diagnostic is appropriate. Using this method, our data show no signs of deviation from normality in $\mathbf{E}^*$.

\begin{figure}[htbp]
    \centering
    \includegraphics[width=0.7\linewidth]{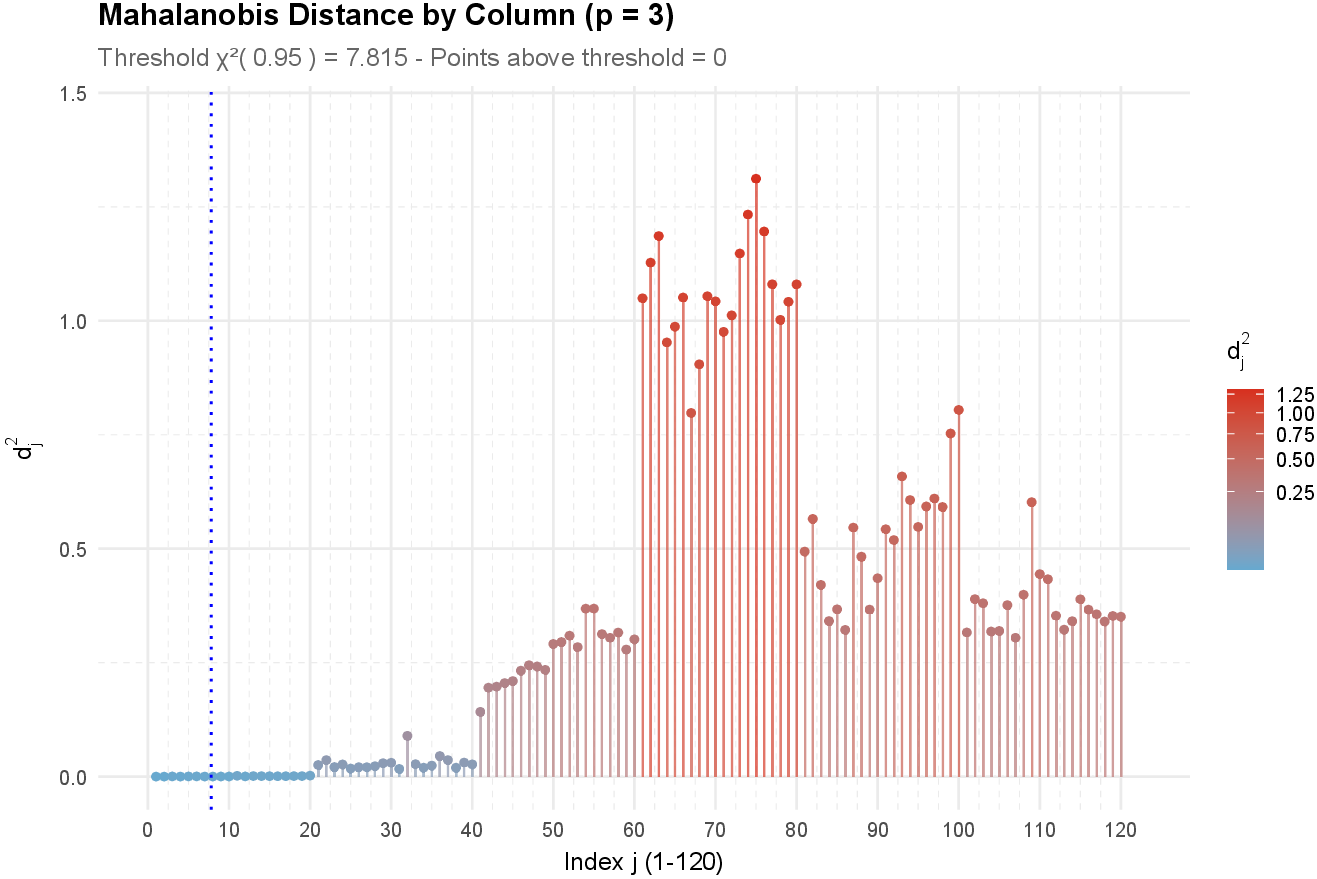}
    \caption{Brazilian agricultural data. Squared Mahalanobis distances ($d_j^2$) for the 120 spatio-temporal observations with $\chi^2_{0.95}(3) = 7.8147$ threshold.}
    \label{fig:mahalanobis_plot}
\end{figure}

We computed the squared Mahalanobis distance $d_j^2$ for each of the 120 spatio-temporal observations (6 cities over 20 years) to check for outliers and validate the multivariate normality assumption. In Figure~\ref{fig:mahalanobis_plot}, each distance appears as a vertical segment colored from blue (low values) to red (high values). All values lie below the $\chi^2_{0.95}(3) = 7.8147$ threshold, indicating no local outliers. This supports the matrix-normal specification, with the concentration of smaller distances in the first half of the plot reflecting the hierarchical spatio-temporal organization.

The row-wise metric $r_i^2$ evaluates each unit's profile across all 120 conditions. The values
\[
[r_1^2, r_2^2, r_3^2] = [15.05912,\ 15.00415,\ 15.00617]
\]
show striking consistency and fall well below the expected $\chi^2_{120}$ mean, indicating homogeneous patterns across units with no systematically aberrant profiles.

\begin{figure}[!ht]
\centering
\includegraphics[width=0.7\textwidth]{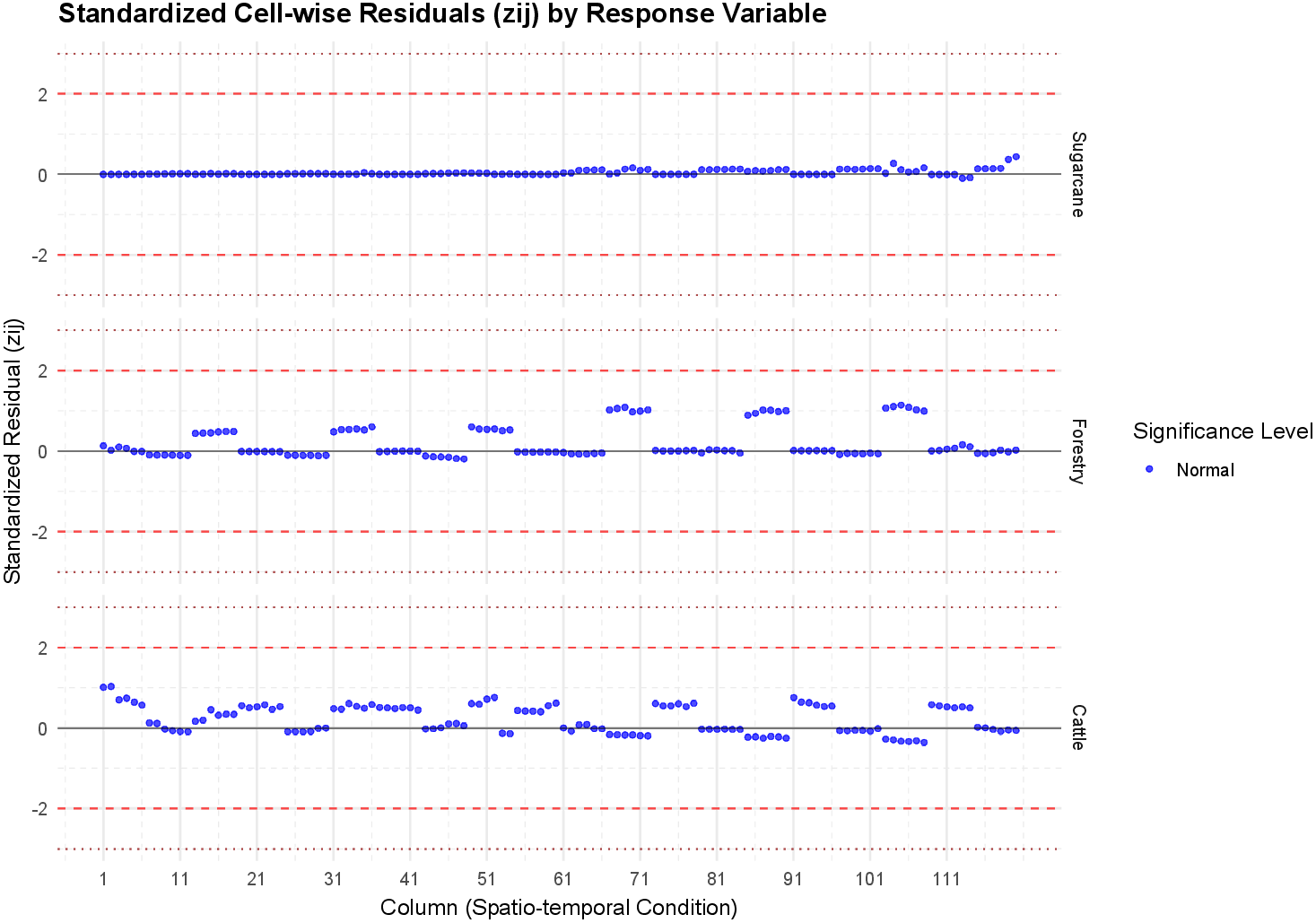}
\caption{Brazilian agricultural data. Standardized cell-wise residuals \(z_{ij}\) for the three response variables, showing individual deviations after accounting for covariance structures.}
\label{fig:cellwise_residuals}
\end{figure}

The cell-wise residuals \(z_{ij}\) in Figure~\ref{fig:cellwise_residuals} mostly cluster near zero within the expected $[-2,2]$ range, indicating a good local fit. No systematic patterns or outlier clusters emerge across the 120 conditions, though forestry and cattle show slightly larger deviations. With no values exceeding the 95\% confidence bounds, the matrix-normal specification appears adequate even at this granular level.

The parameter estimates for the selected Model 2 provide a clear interpretation of the covariate effects and model structure. The coefficient matrix \(\widehat{\mathbf{B}}\) reveals that the first covariate (harvested area of temporary crops) is the dominant driver, exhibiting a strong positive effect on sugarcane yield (\(\beta_{11} = 0.093283\)), while its associations with forestry and cattle are more subtle (\(\beta_{21} = 0.005035\) and \(\beta_{31} = -0.001910\), respectively). The second covariate (permanent crops area) shows modest positive effects on sugarcane and cattle (\(\beta_{12} = 0.007515\), \(\beta_{32} = 0.000980\)) and a negligible impact on forestry (\(\beta_{22} = 0.000194\)). Crucially, the constraint \(\beta_{23} = 0\) is imposed, meaning the third covariate (GDP, \(X_3\)) has no linear effect on the second response (forestry, \(Y_2\)) across all cities and years, a specification informed by our initial exploratory analysis that found this relationship to be statistically insignificant.

The estimated row covariance matrix \(\widehat{\boldsymbol{\Sigma}}\), which models the dependence between the response variables, confirms very weak residual dependence among them after accounting for the model, with all off-diagonal elements having an absolute value within 0.012. For instance, the residual correlations between sugarcane and cattle, sugarcane and forestry, and forestry and cattle are approximately 0.0116, -0.0036, and nearly zero, respectively.  The residual variances, \(\text{diag}(\widehat{\boldsymbol{\Sigma}}) = (1.000000, 0.001665, 0.053587)\), are highly heterogeneous, indicating that sugarcane has the highest unexplained variability, cattle has moderate unexplained variability, and forestry has very little. Furthermore, the scalar parameters indicate a spatio-temporal structure characterized by moderate spatial dependence (\(\phi_s = 1.456300\)) and very strong temporal persistence (\(\rho = 0.992700\)), with a spatial variance of \(\sigma^2 = 7.995700\). The spatial decay parameter \(\phi_s\) suggests a broader spatial influence, while the temporal autocorrelation \(\rho\), being very close to 1, points to high year-to-year consistency. Collectively, these estimates support the adequacy of Model 2 in capturing the essential dynamics of the data, characterized by high temporal continuity and moderate spatial heterogeneity across municipalities.

\begin{table}[H]
\centering
\footnotesize
\caption{Parameter estimates for the matrix-variate regression model for Brazilian agricultural data.}
\label{parameter_est}
\begin{tabular}{cc}
\begin{tabular}[t]{l@{}c}
\toprule
\textbf{Model 1} & \\
\midrule
\multicolumn{2}{l}{\textbf{Matrix}} \\
\addlinespace
$\boldsymbol{B}$ &
$\begin{bmatrix}
  0.093079 & 0.007534 & 0.055047 \\ 
 0.005086 & 0.000196 & 0.000442 \\ 
-0.001900 & 0.000980 & 0.003399 \\
\end{bmatrix}$ \\
\addlinespace
$\boldsymbol{\Sigma}$ &
$\begin{bmatrix}
 1.000000 & -0.003610 & 0.011604 \\ 
-0.003610 &  0.001669 & 0.000004 \\ 
 0.011604 &  0.000004 & 0.053560 \\ 
\end{bmatrix}$ \\
\midrule
\multicolumn{2}{l}{\textbf{Scalar Parameters}} \\
\addlinespace
$\phi_s$       & 1.456300 \\
$\rho$       & 0.992500 \\
$\sigma^{2}$ & 7.765000 \\
BIC & -691.091400\\
\bottomrule
\end{tabular}
&
\begin{tabular}[t]{l@{}c}
\toprule
\textbf{Model 2} & \\
\midrule
\multicolumn{2}{l}{\textbf{Matrix}} \\
\addlinespace
$\boldsymbol{B}$ &
$\begin{bmatrix}
0.093283 & 0.007515 & 0.055092 \\ 
 0.005035 & 0.000194 & 0.000000 \\ 
-0.001910 & 0.000980 & 0.003403 \\ 
\end{bmatrix}$ \\
\addlinespace
$\boldsymbol{\Sigma}$ &
$\begin{bmatrix}
 1.000000 & -0.003604 & 0.011590 \\ 
-0.003604 &  0.001665 & 0.000003 \\ 
 0.011590 &  0.000003 & 0.053587 \\ 
\end{bmatrix}$ \\
\midrule
\multicolumn{2}{l}{\textbf{Scalar Parameters}} \\
\addlinespace
$\phi_s$       & 1.456300 \\
$\rho$       & 0.992700 \\
$\sigma^{2}$ & 7.995700 \\
BIC & -696.824300\\
\bottomrule
\end{tabular}
\end{tabular}
\end{table}

In summary, the model provides a coherent representation of the data. The absence of outliers across all diagnostic metrics indicates that the specification, including the covariance structures $\boldsymbol{\Sigma}$ and $\boldsymbol{\Psi}$, adequately captures the data variability. Standardized residuals align with the matrix-normal assumption, and parameter estimates appear robust, free of distortion from influential points.

Substantively, one covariate (likely planted area) dominates sugarcane production, while weak residual correlations suggest that the model captures most of the systematic variation. The very high temporal autocorrelation ($\rho = 0.9927$) shows strong persistence over time, while the spatial dependence ($\phi_s = 1.4563$) indicates localized effects. Overall, the model effectively uncovers meaningful patterns in multivariate agricultural data across space and time.

\section{Conclusion}\label{sec:conclusion}.

This paper introduced a matrix-variate regression model specifically designed to analyze multivariate response data indexed over spatial locations and time points. The model accommodates dependencies among response variables through a row covariance matrix and incorporates spatial and temporal dependence in the columns via a Kronecker-structured covariance matrix. By organizing the data in matrix form, the proposed approach enables joint modeling of cross-variable, spatial, and temporal correlations.

A key feature of the model is the flexibility in specifying the structure of the coefficient matrix $\mathbf{B}$, which links covariates to the multivariate response. We explored several configurations, including identity, diagonal, full, sparse, and block structures, as well as formulations incorporating interaction terms and polynomial expansions. These alternatives enable the model to adapt to diverse analytical goals while balancing interpretability and computational feasibility.

We derived ML estimators for all model parameters and presented detailed procedures for their computation. A simulation study demonstrated the model's ability to recover the underlying parameters across scenarios with varying spatial and temporal resolutions. The estimation procedures showed satisfactory performance for moderately large samples, although it genuinely high-dimensional settings (e.g., with many response variables or covariates) were not explored.

The real-data application involving agricultural and livestock production across Brazilian municipalities further illustrated the model's capability to capture structured dependence across space and time. Sugarcane production exhibited strong temporal persistence and high residual variability, while cattle and forestry responses were more stable across locations and over time. The analysis also underscored the critical role of covariates, particularly planted area, in explaining variation in the multivariate responses.

The residual analysis, conducted using Mahalanobis distances ($d_j^2$), row-wise metrics ($r_i^2$), and cell-wise residuals ($z_{ij}$), provides a comprehensive diagnostic assessment of the matrix-normal model. While systematic diagnostic frameworks for matrix-variate regression remain an area for further methodological development, the approaches applied here successfully validated model assumptions, detected no outliers, and confirmed the adequacy of the chosen covariance structures. Expanding these diagnostics, particularly for influence analysis and goodness-of-fit measures, represents a valuable direction for future methodological work in matrix-variate settings.

While the model offers a coherent and structured framework for spatio-temporal data analysis, certain limitations must be acknowledged. The assumption of a separable Kronecker structure, while simplifying computation, may be restrictive in more complex applications. Additionally, reliance on numerical optimization and matrix operations can pose challenges for large-scale problems. Future research may consider regularization techniques, Bayesian formulations, and extensions to accommodate non-Gaussian responses or dynamic covariance structures.

\section*{Acknowledgments}

Victor H. Lachos acknowledges the partial financial support from the Office of the Vice President for Research (OVPR) Research Excellence Program (REP).  Carlos A. R. Diniz gratefully acknowledges the financial support provided by FAPESP (São Paulo Research Foundation), Grant No. 2024/08369-8.

\section*{Appendix}

This appendix presents the complete dataset used in the empirical application of Section 4.

\begin{scriptsize}
\begin{longtable}{l|ccc|ccc}
\caption{Brazilian agricultural data. Key agricultural metrics (Y) and agricultural covariates (X) for six municipalities in Minas Gerais, Brazil. Y in millions, X in thousands, GDP in billions.}
\label{tab:YXdata} \\
\hline
Municipality & Y1 & Y2 & Y3 & X1 & X2 & X3 \\
\hline
\endfirsthead
\multicolumn{7}{c}{\tablename\ \thetable\ -- Continued from previous page} \\
\hline
Municipality & Y1 & Y2 & Y3 & X1 & X2 & X3 \\
\hline
\endhead
\hline \multicolumn{7}{r}{\textit{Continued on next page}} \\
\endfoot
\hline
\endlastfoot
\multicolumn{4}{c}{$\mathbf{Y}_{2002}$} & \multicolumn{3}{c}{$\mathbf{X}_{2002}$} \\
\hline
& Sugarcane & Forestry & Cattles & Temp crops & Perm Crops & GDP \\
\hline
Gonzaga (MG) & 0.002000 & 0.000298 & 0.003890 & 0.504 & 0.225 & 0.009909 \\
Guanhães (MG) & 0.015000 & 0.237825 & 0.032765 & 1.505 & 0.207 & 0.115792 \\
Patos de Minas (MG) & 0.010800 & 0.000692 & 0.158038 & 18.789 & 6.616 & 0.731658 \\
Prata (MG) & 0.037848 & 0.034260 & 0.348400 & 4.152 & 2.780 & 0.145882 \\
Uberaba (MG) & 0.823230 & 0.050030 & 0.258896 & 96.285 & 2.264 & 2.862217 \\
Uberlândia (MG) & 0.003000 & 0.029880 & 0.209757 & 52.727 & 3.570 & 5.386240 \\
\hline
\multicolumn{4}{c}{$\mathbf{Y}_{2003}$} & \multicolumn{3}{c}{$\mathbf{X}_{2003}$} \\
\hline
& Sugarcane & Forestry & Cattles & Temp crops & Perm Crops & GDP \\
\hline
Gonzaga (MG) & 0.002000 & 0.001957 & 0.002835 & 0.500 & 0.225 & 0.009906 \\
Guanhães (MG) & 0.019250 & 0.299295 & 0.034617 & 1.524 & 0.215 & 0.142073 \\
Patos de Minas (MG) & 0.008100 & 0.000752 & 0.173250 & 19.232 & 5.196 & 0.863434 \\
Prata (MG) & 0.036000 & 0.036316 & 0.369304 & 28.108 & 3.496 & 0.192014 \\
Uberaba (MG) & 1.530000 & 0.005820 & 0.276847 & 125.671 & 2.024 & 3.526804 \\
Uberlândia (MG) & 0.003000 & 0.031374 & 0.231526 & 53.885 & 3.881 & 6.467993 \\
\hline
\multicolumn{4}{c}{$\mathbf{Y}_{2004}$} & \multicolumn{3}{c}{$\mathbf{X}_{2004}$} \\
\hline
& Sugarcane & Forestry & Cattles & Temp crops & Perm Crops & GDP \\
\hline
Gonzaga (MG) & 0.002000 & 0.032244 & 0.004557 & 0.299 & 0.185 & 0.012350 \\
Guanhães (MG) & 0.022800 & 0.182550 & 0.036262 & 1.582 & 0.170 & 0.150544 \\
Patos de Minas (MG) & 0.015300 & 0.000780 & 0.177070 & 19.247 & 5.844 & 1.043779 \\
Prata (MG) & 0.038400 & 0.039584 & 0.379869 & 34.081 & 3.598 & 0.256913 \\
Uberaba (MG) & 1.441055 & 0.000110 & 0.249168 & 150.352 & 2.159 & 3.894232 \\
Uberlândia (MG) & 0.003000 & 0.028029 & 0.231352 & 66.132 & 4.078 & 7.313188 \\
\hline
\multicolumn{4}{c}{$\mathbf{Y}_{2005}$} & \multicolumn{3}{c}{$\mathbf{X}_{2005}$} \\
\hline
& Sugarcane & Forestry & Cattles & Temp crops & Perm Crops & GDP \\
\hline
Gonzaga (MG) & 0.004000 & 0.004447 & 0.004521 & 0.360 & 0.081 & 0.013175 \\
Guanhães (MG) & 0.022800 & 0.225898 & 0.037032 & 1.417 & 0.175 & 0.172885 \\
Patos de Minas (MG) & 0.015480 & 0.000826 & 0.180803 & 20.746 & 5.978 & 1.126453 \\
Prata (MG) & 0.032000 & 0.040948 & 0.342089 & 34.406 & 3.396 & 0.277033 \\
Uberaba (MG) & 1.900000 & 0.000120 & 0.230557 & 165.953 & 2.417 & 3.843969 \\
Uberlândia (MG) & 0.004500 & 0.029991 & 0.213485 & 70.680 & 3.979 & 7.887835 \\
\hline
\multicolumn{4}{c}{$\mathbf{Y}_{2006}$} & \multicolumn{3}{c}{$\mathbf{X}_{2006}$} \\
\hline
& Sugarcane & Forestry & Cattles & Temp crops & Perm Crops & GDP \\
\hline
Gonzaga (MG) & 0.004000 & 0.029075 & 0.004981 & 0.440 & 0.081 & 0.016805 \\
Guanhães (MG) & 0.024000 & 0.132358 & 0.037981 & 1.318 & 0.163 & 0.173690 \\
Patos de Minas (MG) & 0.015300 & 0.000061 & 0.182127 & 19.796 & 5.954 & 1.240373 \\
Prata (MG) & 0.036000 & 0.040464 & 0.344664 & 18.839 & 3.396 & 0.292518 \\
Uberaba (MG) & 3.060000 & 0.001410 & 0.226136 & 161.004 & 2.733 & 4.286142 \\
Uberlândia (MG) & 0.004500 & 0.031191 & 0.213210 & 68.210 & 4.058 & 9.233092 \\
\hline
\multicolumn{4}{c}{$\mathbf{Y}_{2007}$} & \multicolumn{3}{c}{$\mathbf{X}_{2007}$} \\
\hline
& Sugarcane & Forestry & Cattles & Temp crops & Perm Crops & GDP \\
\hline
Gonzaga (MG) & 0.003000 & 0.029621 & 0.005525 & 0.355 & 0.070 & 0.017784 \\
Guanhães (MG) & 0.025200 & 0.169072 & 0.037657 & 1.307 & 0.153 & 0.201090 \\
Patos de Minas (MG) & 0.015300 & 0.000067 & 0.190687 & 19.471 & 6.086 & 1.364968 \\
Prata (MG) & 0.066400 & 0.038979 & 0.352984 & 8.537 & 3.438 & 0.356423 \\
Uberaba (MG) & 3.315000 & 0.002502 & 0.209812 & 158.147 & 2.938 & 5.377202 \\
Uberlândia (MG) & 0.004500 & 0.036805 & 0.228565 & 61.494 & 4.058 & 9.653788 \\
\hline
\multicolumn{4}{c}{$\mathbf{Y}_{2008}$} & \multicolumn{3}{c}{$\mathbf{X}_{2008}$} \\
\hline
& Sugarcane & Forestry & Cattles & Temp crops & Perm Crops & GDP \\
\hline
Gonzaga (MG) & 0.003000 & 0.003470 & 0.005214 & 0.345 & 0.073 & 0.018799 \\
Guanhães (MG) & 0.025800 & 0.148712 & 0.039943 & 1.186 & 0.128 & 0.230212 \\
Patos de Minas (MG) & 0.015300 & 0.000071 & 0.195422 & 23.785 & 6.101 & 1.538473 \\
Prata (MG) & 0.066400 & 0.037344 & 0.308602 & 7.797 & 3.188 & 0.384859 \\
Uberaba (MG) & 5.467500 & 0.003593 & 0.218008 & 161.251 & 2.907 & 6.234095 \\
Uberlândia (MG) & 0.028500 & 0.036800 & 0.205709 & 66.174 & 3.845 & 12.666848 \\
\hline
\multicolumn{4}{c}{$\mathbf{Y}_{2009}$} & \multicolumn{3}{c}{$\mathbf{X}_{2009}$} \\
\hline
& Sugarcane & Forestry & Cattles & Temp crops & Perm Crops & GDP \\
\hline
Gonzaga (MG) & 0.003000 & 0.000692 & 0.004736 & 0.345 & 0.073 & 0.021018 \\
Guanhães (MG) & 0.031500 & 0.169410 & 0.040253 & 1.150 & 0.107 & 0.238306 \\
Patos de Minas (MG) & 0.042300 & 0.000079 & 0.195346 & 24.834 & 6.226 & 1.720005 \\
Prata (MG) & 0.720000 & 0.036405 & 0.326462 & 13.792 & 3.188 & 0.455824 \\
Uberaba (MG) & 4.227500 & 0.004685 & 0.229600 & 149.331 & 2.341 & 6.450398 \\
Uberlândia (MG) & 0.081000 & 0.040160 & 0.232409 & 64.487 & 3.845 & 14.602122 \\
\hline
\multicolumn{4}{c}{$\mathbf{Y}_{2010}$} & \multicolumn{3}{c}{$\mathbf{X}_{2010}$} \\
\hline
& Sugarcane & Forestry & Cattles & Temp crops & Perm Crops & GDP \\
\hline
Gonzaga (MG) & 0.003000 & 0.012186 & 0.005363 & 0.345 & 0.073 & 0.023936 \\
Guanhães (MG) & 0.032200 & 0.211399 & 0.042859 & 1.106 & 0.101 & 0.314400 \\
Patos de Minas (MG) & 0.081990 & 0.000114 & 0.192688 & 24.751 & 6.179 & 1.996955 \\
Prata (MG) & 0.720000 & 0.037012 & 0.352579 & 17.576 & 3.309 & 0.541173 \\
Uberaba (MG) & 4.370000 & 0.137644 & 0.189097 & 135.423 & 2.192 & 7.299720 \\
Uberlândia (MG) & 0.072000 & 0.000165 & 0.278890 & 65.400 & 3.845 & 18.950577 \\
\hline
\multicolumn{4}{c}{$\mathbf{Y}_{2011}$} & \multicolumn{3}{c}{$\mathbf{X}_{2011}$} \\
\hline
& Sugarcane & Forestry & Cattles & Temp crops & Perm Crops & GDP \\
\hline
Gonzaga (MG) & 0.000800 & 0.000914 & 0.005875 & 0.355 & 0.073 & 0.030601 \\
Guanhães (MG) & 0.033600 & 0.220123 & 0.042644 & 1.185 & 0.085 & 0.352159 \\
Patos de Minas (MG) & 0.081990 & 0.010409 & 0.210918 & 25.507 & 6.131 & 2.341902 \\
Prata (MG) & 0.880000 & 0.035000 & 0.367300 & 15.205 & 6.563 & 0.621333 \\
Uberaba (MG) & 4.370000 & 0.155000 & 0.211653 & 133.340 & 2.312 & 8.125741 \\
Uberlândia (MG) & 0.048000 & 0.000180 & 0.235000 & 68.037 & 2.398 & 19.553210 \\
\hline
\multicolumn{4}{c}{$\mathbf{Y}_{2012}$} & \multicolumn{3}{c}{$\mathbf{X}_{2012}$} \\
\hline
& Sugarcane & Forestry & Cattles & Temp crops & Perm Crops & GDP \\
\hline
Gonzaga (MG) & 0.001500 & 0.072180 & 0.006057 & 0.210 & 0.150 & 0.035022 \\
Guanhães (MG) & 0.033600 & 0.048132 & 0.042000 & 0.735 & 0.060 & 0.455001 \\
Patos de Minas (MG) & 0.081990 & 0.020705 & 0.212987 & 22.586 & 6.414 & 2.655514 \\
Prata (MG) & 0.880000 & 0.025000 & 0.356491 & 14.890 & 6.572 & 0.672546 \\
Uberaba (MG) & 5.700000 & 0.126000 & 0.207700 & 144.866 & 2.605 & 9.399680 \\
Uberlândia (MG) & 0.048000 & 0.037905 & 0.231313 & 74.656 & 2.492 & 22.837278 \\
\hline
\multicolumn{4}{c}{$\mathbf{Y}_{2013}$} & \multicolumn{3}{c}{$\mathbf{X}_{2013}$} \\
\hline
& Sugarcane & Forestry & Cattles & Temp crops & Perm Crops & GDP \\
\hline
Gonzaga (MG) & 0.001500 & 0.028682 & 0.005988 & 0.320 & 0.150 & 0.036794 \\
Guanhães (MG) & 0.033600 & 0.502642 & 0.042574 & 1.083 & 0.060 & 0.503618 \\
Patos de Minas (MG) & 0.081990 & 0.031000 & 0.217469 & 23.795 & 6.419 & 2.175491 \\
Prata (MG) & 0.600000 & 0.040250 & 0.371550 & 24.190 & 7.760 & 0.748056 \\
Uberaba (MG) & 5.700000 & 0.139887 & 0.204801 & 146.857 & 2.590 & 10.864034 \\
Uberlândia (MG) & 0.740000 & 0.044890 & 0.223000 & 79.670 & 4.319 & 25.718586 \\
\hline
\multicolumn{4}{c}{$\mathbf{Y}_{2014}$} & \multicolumn{3}{c}{$\mathbf{X}_{2014}$} \\
\hline
& Sugarcane & Forestry & Cattles & Temp crops & Perm Crops & GDP \\
\hline
Gonzaga (MG) & 0.001500 & 0.058031 & 0.006230 & 0.320 & 0.150 & 0.041765 \\
Guanhães (MG) & 0.029250 & 0.210276 & 0.040200 & 1.083 & 0.060 & 0.542735 \\
Patos de Minas (MG) & 0.057150 & 0.030000 & 0.209512 & 24.536 & 6.421 & 3.510975 \\
Prata (MG) & 0.338400 & 0.032500 & 0.393870 & 27.534 & 7.770 & 0.812242 \\
Uberaba (MG) & 6.885000 & 0.197120 & 0.200460 & 153.445 & 2.630 & 11.564234 \\
Uberlândia (MG) & 0.850400 & 0.028000 & 0.213800 & 86.315 & 4.453 & 28.390937 \\
\hline
\multicolumn{4}{c}{$\mathbf{Y}_{2015}$} & \multicolumn{3}{c}{$\mathbf{X}_{2015}$} \\
\hline
& Sugarcane & Forestry & Cattles & Temp crops & Perm Crops & GDP \\
\hline
Gonzaga (MG) & 0.004800 & 0.048726 & 0.006340 & 0.135 & 0.090 & 0.042237 \\
Guanhães (MG) & 0.017500 & 0.103652 & 0.042300 & 0.482 & 0.012 & 0.584927 \\
Patos de Minas (MG) & 0.071820 & 0.028400 & 0.228960 & 24.996 & 5.315 & 3.810486 \\
Prata (MG) & 0.687520 & 0.038750 & 0.414700 & 21.616 & 9.656 & 0.824339 \\
Uberaba (MG) & 6.266200 & 0.371158 & 0.203475 & 147.466 & 1.773 & 12.514706 \\
Uberlândia (MG) & 0.682720 & 0.148250 & 0.217560 & 85.875 & 4.418 & 29.472293 \\
\hline
\multicolumn{4}{c}{$\mathbf{Y}_{2016}$} & \multicolumn{3}{c}{$\mathbf{X}_{2016}$} \\
\hline
& Sugarcane & Forestry & Cattles & Temp crops & Perm Crops & GDP \\
\hline
Gonzaga (MG) & 0.005000 & 0.000320 & 0.006380 & 0.260 & 0.083 & 0.045948 \\
Guanhães (MG) & 0.026000 & 0.124704 & 0.046251 & 0.730 & 0.012 & 0.608869 \\
Patos de Minas (MG) & 0.071820 & 0.002650 & 0.234092 & 28.126 & 6.325 & 4.195265 \\
Prata (MG) & 0.687520 & 0.103160 & 0.426852 & 21.304 & 9.887 & 0.946465 \\
Uberaba (MG) & 6.267158 & 0.439160 & 0.208813 & 194.020 & 1.505 & 13.463502 \\
Uberlândia (MG) & 0.682720 & 0.226906 & 0.229797 & 86.380 & 4.668 & 32.553439 \\
\hline
\multicolumn{4}{c}{$\mathbf{Y}_{2017}$} & \multicolumn{3}{c}{$\mathbf{X}_{2017}$} \\
\hline
& Sugarcane & Forestry & Cattles & Temp crops & Perm Crops & GDP \\
\hline
Gonzaga (MG) & 0.000210 & 0.003996 & 0.009789 & 0.143 & 0.015 & 0.048236 \\
Guanhães (MG) & 0.002420 & 0.301348 & 0.045304 & 0.315 & 0.017 & 0.617677 \\
Patos de Minas (MG) & 0.017100 & 0.006120 & 0.219002 & 41.815 & 6.153 & 4.378909 \\
Prata (MG) & 0.617696 & 0.152225 & 0.407810 & 24.487 & 9.405 & 0.928682 \\
Uberaba (MG) & 6.768000 & 0.309343 & 0.185260 & 164.466 & 1.476 & 13.153000 \\
Uberlândia (MG) & 0.716256 & 0.205340 & 0.224450 & 90.476 & 5.243 & 34.211313 \\
\hline
\multicolumn{4}{c}{$\mathbf{Y}_{2018}$} & \multicolumn{3}{c}{$\mathbf{X}_{2018}$} \\
\hline
& Sugarcane & Forestry & Cattles & Temp crops & Perm Crops & GDP \\
\hline
Gonzaga (MG) & 0.000224 & 0.018430 & 0.010355 & 0.188 & 0.021 & 0.051196 \\
Guanhães (MG) & 0.002600 & 0.249511 & 0.044598 & 0.316 & 0.026 & 0.694601 \\
Patos de Minas (MG) & 0.017000 & 0.006050 & 0.217062 & 41.494 & 6.311 & 4.697187 \\
Prata (MG) & 0.850000 & 0.295950 & 0.393879 & 22.323 & 10.850 & 0.977983 \\
Uberaba (MG) & 6.800000 & 0.185000 & 0.176600 & 134.060 & 1.549 & 14.732211 \\
Uberlândia (MG) & 0.635000 & 0.139001 & 0.216038 & 91.165 & 5.352 & 37.513607 \\
\hline
\multicolumn{4}{c}{$\mathbf{Y}_{2019}$} & \multicolumn{3}{c}{$\mathbf{X}_{2019}$} \\
\hline
& Sugarcane & Forestry & Cattles & Temp crops & Perm Crops & GDP \\
\hline
Gonzaga (MG) & 0.000480 & 0.016642 & 0.010819 & 0.178 & 0.021 & 0.054420 \\
Guanhães (MG) & 0.002631 & 0.046238 & 0.045266 & 0.398 & 0.026 & 0.786211 \\
Patos de Minas (MG) & 0.016826 & 0.015300 & 0.221113 & 46.215 & 6.313 & 4.957415 \\
Prata (MG) & 0.842363 & 0.206600 & 0.387956 & 21.063 & 11.798 & 1.002600 \\
Uberaba (MG) & 6.800000 & 0.045000 & 0.173497 & 140.506 & 1.598 & 15.545269 \\
Uberlândia (MG) & 0.635000 & 0.058630 & 0.210520 & 92.181 & 5.782 & 37.638742 \\
\hline
\multicolumn{4}{c}{$\mathbf{Y}_{2020}$} & \multicolumn{3}{c}{$\mathbf{X}_{2020}$} \\
\hline
& Sugarcane & Forestry & Cattles & Temp crops & Perm Crops & GDP \\
\hline
Gonzaga (MG) & 0.000627 & 0.060413 & 0.011006 & 0.190 & 0.021 & 0.060207 \\
Guanhães (MG) & 0.003531 & 0.198145 & 0.046044 & 0.345 & 0.026 & 0.793589 \\
Patos de Minas (MG) & 0.017670 & 0.014100 & 0.220580 & 45.164 & 9.070 & 5.401454 \\
Prata (MG) & 0.948600 & 0.239000 & 0.395112 & 21.346 & 12.083 & 1.135176 \\
Uberaba (MG) & 7.975320 & 0.345000 & 0.173967 & 149.757 & 1.658 & 17.190845 \\
Uberlândia (MG) & 0.736960 & 0.115500 & 0.217714 & 88.911 & 5.693 & 37.631537 \\
\hline
\multicolumn{4}{c}{$\mathbf{Y}_{2021}$} & \multicolumn{3}{c}{$\mathbf{X}_{2021}$} \\
\hline
& Sugarcane & Forestry & Cattles & Temp crops & Perm Crops & GDP \\
\hline
Gonzaga (MG) & 0.000512 & 0.000000 & 0.011220 & 0.203 & 0.019 & 0.068973 \\
Guanhães (MG) & 0.002880 & 0.140344 & 0.048013 & 0.366 & 0.025 & 0.996965 \\
Patos de Minas (MG) & 0.012750 & 0.013200 & 0.227968 & 47.123 & 9.209 & 6.145130 \\
Prata (MG) & 0.912000 & 0.217000 & 0.400765 & 25.470 & 11.718 & 1.293341 \\
Uberaba (MG) & 8.613500 & 0.389000 & 0.172501 & 168.519 & 1.551 & 20.397519 \\
Uberlândia (MG) & 0.698400 & 0.112000 & 0.208070 & 89.334 & 5.538 & 43.129285 \\
\hline
\end{longtable}
\end{scriptsize}

\bibliographystyle{plainnat}
\bibliography{refs}               
\end{document}